\documentclass[aip,pop,preprint,showpacs,notocite,superscriptaddress]{revtex4-1}
\usepackage{blindtext}
\usepackage{amssymb, amsmath,framed}
	\usepackage{mathrsfs}
\usepackage{bm}
\usepackage[colorlinks=true,linkcolor=blue]{hyperref}
\usepackage{amsfonts,amstext,graphics}
\usepackage{subfigure}
\usepackage{graphicx,epstopdf}
\usepackage{xcolor}
	\hypersetup
	{
		colorlinks,%
		citecolor=red,%
		linkcolor=blue,%
		urlcolor=green,%
	}
\expandafter\ifx\csname package@font\endcsname\relax\else
 \expandafter\expandafter
 \expandafter\usepackage
 \expandafter\expandafter
 \expandafter{\csname package@font\endcsname}
\fi
\newcommand{\bv}{\mathbf{V}}

\newcommand{\de}{d_{e}^{2}}
\newcommand{\hz}{\hat{z}}
\newcommand{\pe}{\psi^{*}}
\newcommand{\lap}{\nabla^2}

\newcommand{\mcal}{\mathcal{L}}

\newcommand{\ben}{\begin{eqnarray}}
\newcommand{\een}{\end{eqnarray}}
\newcommand{\beq}{\begin{equation}}
\newcommand{\eeq}{\end{equation}}

\newcommand{\pa}{\partial}
\newcommand{\curl}{\nabla\times}
\newcommand{\pl}{\psi_{\lambda_{\pm}}}
\newcommand{\bl}{b_{\lambda_{\pm}}}
\newcommand{\bbl}{\mathbf{B}_{\lambda_{\pm}}}

\newcommand{\ccal}{\mathcal{C}}

\begin{document}

\title{Structure and computation of two-dimensional incompressible extended MHD}

\author{D.Grasso}
\affiliation{ISC-CNR and Politecnico di Torino, Dipartimento di Energia, C.so Duca degli Abruzzi 24, 10129 Torino, Italy }
\email{daniela.grasso@infm.polito.it}
\author{E. Tassi}
\affiliation{Aix Marseille Univ, Universit\'e de Toulon, CNRS, CPT, Marseille, France}
\email{tassi@cpt.univ-mrs.fr}
\author{ H. M. Abdelhamid}
\affiliation{Graduate School of Frontier Sciences, The University of Tokyo, Kashiwanoha, Kashiwa, Chiba 277-8561, Japan}
\affiliation{Physics Department, Faculty of Science, Mansoura University, Mansoura 35516, Egypt}
\email{hamdi@ppl.k.u-tokyo.ac.jp}
 \author{P. J. Morrison}
 \affiliation{Department of Physics and Institute for Fusion Studies, The University of Texas at Austin}
 \email{morrison@physics.utexas.edu}

\begin{abstract}
A comprehensive study of a reduced version of L\"ust's equations, the extended magnetohydrodynamic (XMHD) model obtained  from the two-fluid theory for electrons and ions with the enforcement of quasineutrality, is given.   Starting from the Hamiltonian structure of the fully three-dimensional theory, a Hamiltonian two-dimensional incompressible four-field model is derived.  In this way energy conservation along with four families of Casimir invariants are naturally obtained.  The construction facilitates various limits leading to the Hamiltonian  forms of Hall, inertial, and ideal MHD, with their conserved energies and Casimir invariants.  Basic linear theory of the four-field model  is treated, and the growth rate for collisionless reconnection is obtained.  Results from nonlinear simulations of collisionless tearing are presented and interpreted using, in particular normal fields,   a product of the Hamiltonian theory that gives rise to simplified equations of motion.

\end{abstract}
\maketitle

%\tableofcontents

%%%%%%%%%%%%%%%%%%%%%%%%%%%%%%%
%%%%%%%%%%%%%%%%%%%%%%%%%%%%%%%
%%%%%%%%%%%%%%%%%%%%%%%%%%%%%%%
\section{Introduction} \label{sec:intro}

Out of necessity for practicable computation, for many decades researchers have produced reduced fluid models for describing aspects of laboratory and naturally occurring plasmas. Applications of such models include  the  exploration of  MHD kink modes by means of reduced MHD,\cite{rmhd} subsequently extended to include, among other effects, Hall and gyro physics \cite{HKM85} as well as parallel compressibility and diamagnetic effects.\cite{DA84}  Further applications range  from the investigation of drift-waves \cite{HW83} to low-frequency turbulence \cite{Bri92,Dor93,Sny01,SSB05} and magnetic reconnection. \cite{SPK94,GCPP01,Grasso07,TMWG08,WHM09,Sco10,Wae12,KWM15} 

 The various models have been obtained  by various means: in some cases rigorous asymptotics were employed, while other models were built on intuition,  or using the device of effecting closure by  constraining to match a desired linear theory (e.g.\ Ref.~\onlinecite{HKM85}).  Based on the noncanonical Hamiltonian formalism introduced for MHD in Ref.~\onlinecite{MG80}  (see e.g.\  Refs.~\onlinecite{M98,M05}  for review) it was advocated in a series of   papers\cite{MH84,MM84,HHM86,Haz87} that retention of  Hamiltonian form can serve  as a derivational aide or as a filter for selecting out  good theories in the ideal limit.  By ideal limit,  is meant  the limit of the model where all dissipative terms, such as collisions, Landau damping, and dissipative anomalous transport terms are neglected.  Subsequently there have been many papers by many authors that have adopted this point of view. 

 In the present work we consider the two-dimensional (2D) incompressible reduction of the XMHD model  derived by L\"ust in Ref.~\onlinecite{Lust}.  This model  is simply a reduced case of a two-fluid model in which the charge quasineutrality condition is invoked, the displacement current is ignored, and the smallness of the electron-ion mass  ratio is taken to the first order approximation. The value  of the XMHD model resides in  its ability  to capture and describe the main two-fluid effects, e.g., Hall drift and electron inertia.  Unlike its parent 3D version, \cite{Abd15,Lin15} the 2D incompressible reduction of XMHD (RXMHD) has not yet been explored, however, from the above mentioned Hamiltonian perspective.   The Hamiltonian approach can indeed be particularly fruitful in this context because of  the richness of the Casimir invariants that typically emerge in 2D models.  These invariants, which are  associated with the Hamiltonian structure,  provide information on the dynamics.   Identifying the Hamiltonian structure of RXMHD and providing its Casimir invariants is one of the goals of this paper. A further related issue that we treat, is that of investigating how the conservation laws related to the Casimir invariants in RXMHD, which properly accounts for both ion and electron physics corrections, compare with those of submodels such as ideal reduced, Hall and inertial MHD, where some of these effects are neglected.

 In addition to the investigation of the Hamiltonian structure of the model, 
we also present an application of RXMHD where the Hamiltonian approach plays a role. Given that RXMHD is a model that extends 2D incompressible Hall MHD by properly accounting for electron physics, a natural application for RXMHD is to 2D magnetic reconnection driven by electron inertia. Such magnetic reconnection has already been studied by means of a model very similar to RXMHD in Ref.~\onlinecite{Bis97}.  These authors considered a weakly dissipative model and identified the fundamental mechanisms of two-fluid collisionless reconnection, in particular with regard to the role of the Hall term and of the electron MHD governing the dynamics at scales below the ion skin depth. In the present manuscript we carry out an investigation of purely non-dissipative magnetic reconnection by means of RXMHD, with an approach that  is somewhat complementary to that adopted in Ref.~\onlinecite{Bis97}. 
We provide an analytical expression for the linear growth rate of reconnecting perturbations  and check it against numerical solutions, and we take advantage of the Hamiltonian formulation to compare the evolution of the physical fields,  in terms of which the model was originally formulated, with  {\em normal fields}, an alternative set of variables.  Normal fields  are associated with the Casimir invariants and express a  simpler dynamics. This approach was used  in previous studies of collisionless reconnection in  Hamiltonian models (see, e.g.\  Refs.~\onlinecite{GCPP01,Del06,Gra09,Tas10,Com12}).   Also,  the Hamiltonian formulation provides the correct expression for the total energy, which we  exploit  in order to investigate the redistribution of magnetic energy into different forms.    We also remark that, in a  recent publication,\cite{And14}  a model very similar to RXMHD was adopted to investigate numerical reconnection rates and the conservation of three invariants during reconnection. 

Our  paper is organized as follows.  In Sec.~\ref{sec:derivation},  XMHD is reviewed and  the Hamiltonian form of RXMHD, a four-field model,  is obtained from  that of the full model.\cite{Abd15,Lin15}   Consequences of this reduction are explored in Sec.~\ref{sec:normalCas} where it is shown that in addition to the energy,  the RXMHD system posses four infinite families of invariants,  the Casimir invariants.  In addition,  the  so-called normal fields are obtained and it is observed that the equations of motion take a particularly simple form when expressed in terms of them (compare Eqs.~\eqref{flux}-\eqref{v} to Eqs.~\eqref{psipm}-\eqref{bpm}).   Section ~\ref{sec:limits} treats various limits of RXMHD leading to reduced Hall MHD (RHMHD), reduced inertial MHD (RIMHD) and reduced MHD (RMHD).  Numerical solution of RXMHD is treated in Sec.~\ref{sec:numerics}.  Here the dispersion relation is plotted for the basic modes of the system, the collisionless tearing instability growth rate  is identified and numerically verified,  and nonlinear simulations of collisionless tearing are performed, which reveal how energy migrates from field into flow.   Finally, in Sec.~\ref{sec:conclu}
 we summarize our results and draw conclusions.

%%%%%%%%%%%%%%%%%%%%%%%%%%%%%%%
%%%%%%%%%%%%%%%%%%%%%%%%%%%%%%%
%%%%%%%%%%%%%%%%%%%%%%%%%%%%%%%
\section{Derivation Of reduced extended magnetohydrodynamics}
\label{sec:derivation}

%%%%%%%%%%%%%%%%%%%%%%%%%%%%%%%
%%%%%%%%%%%%%%%%%%%%%%%%%%%%%%%
\subsection{Extended magnetohydrodynamics}
 \label{SecxMHD}

The governing equations of  extended magnetohydrodynamics (XMHD) are  the continuity equation
\begin{eqnarray}
\label{Cont}
\frac{\partial \rho}{\partial t}=-\nabla\cdot\left(\rho\mathbf{V}\right),
\end{eqnarray}
the force law, 
\begin{eqnarray}
\label{Mom}
\rho\left(\frac{\partial\mathbf{V}}{\partial t}  + \left(\mathbf{V}.\nabla\right)\mathbf{V}\right) &=& -\nabla p+\mathbf{J}\times\mathbf{B}\nonumber\\&&-d^{2}_{e}\left(\mathbf{J}\cdot\nabla\right)\frac{\mathbf{J}}{\rho},
\end{eqnarray}
and the generalized Ohm's law
\begin{eqnarray}
\label{Ohm}
\mathbf{E}+\mathbf{V}\times\mathbf{B}&=&-\frac{d_{i}}{\rho}\nabla p_{e}+d_{i}\frac{\mathbf{J}}{\rho}\times\mathbf{B}\nonumber\\&&+d^{2}_{e}\left[ \frac{\partial}{\partial t}\left(\frac{\mathbf{J}}{\rho}\right)+(\mathbf{V}\cdot\nabla)\frac{\mathbf{J}}{\rho}+\left(\frac{\mathbf{J}}{\rho}\cdot\nabla\right)\mathbf{V}\right]\nonumber\\&&-d_{i}d^{2}_{e}\left(\frac{\mathbf{J}}{\rho}\cdot\nabla\right)\frac{\mathbf{J}}{\rho}.
\end{eqnarray}
Here $\rho$ is the total mass density, $\mathbf{V}$ is the center of mass velocity, $\mathbf{B}$ is the magnetic field, $\mathbf{E}$ is the electric field, $\mathbf{J}$ is the current density and $p=p_{i}+p_{e}$ is the total pressure,  with $p_{i}$ being the ion pressure and $p_{e}$  the electron pressure. The system is normalized to the standard Alfv\'enic units with  $d_{e}=c/(\omega_{pe}L)$   and $d_{i}=c/(\omega_{pi}L)$,  corresponding to the normalized electron and ion skin depths, respectively, where $\omega_{pe}$ and $\omega_{pi}$ are the electron and ion plasma frequencies, and $L$ is the system size.  Equations \eqref{Cont}--\eqref{Ohm} are coupled with the pre-Maxwell equations
\begin{eqnarray}
\label{Max}
\nabla\times\mathbf{E}=-\frac{\partial\mathbf{B}}{\partial t}\qquad   \textrm{and}\qquad  \nabla\times\mathbf{B}=\mathbf{J}, 
\end{eqnarray}
and for this paper the systems will be closed by assuming a barotropic equation of state, i.e., the pressure $p$ is assumed to depend only on the density $\rho$.

Upon using $\rho^{-1}\nabla p=\nabla h\left(\rho\right)$,  which follows from the barotropic assumption, where $h\left(\rho\right)$ is the enthalpy,  and using the pre-Maxwell equations \eqref{Max}, one can obtain, from Eqs.~\eqref{Mom} and \eqref{Ohm}, the following system: 
\begin{eqnarray}
\label{Mom2}
\frac{\partial \mathbf{V}}{\partial t}&=&-\left(\nabla\times\mathbf{V}\right)\times\mathbf{V}+\rho^{-1} \left(\nabla\times\mathbf{B}\right)\times\mathbf{B}^{\ast}\nonumber \\
& &~~~~-\nabla\left(h+V^{2}/2+d^{2}_{e}\left(\nabla\times\mathbf{B}\right)^{2}/2\rho^{2}\right),
\\
\label{Ohm2}
\frac{\partial \bm{B}^{\ast}}{\partial t}&=&\nabla\times\left(\mathbf{V}\times\mathbf{B}^{\ast}\right)- \nabla\times\left(\rho^{-1} \left(\nabla\times\mathbf{B}\right)\times\mathbf{B}^{\ast}\right)\nonumber \\
& &~~~~+d^{2}_{e} \nabla\times\left(\rho^{-1} \left(\nabla\times\mathbf{B}\right)\times\left(\nabla\times\mathbf{V}\right)\right),
\end{eqnarray}
where
\begin{eqnarray}
\label{Bstar}
\mathbf{B}^{\ast}&=&\mathbf{B}+d^{2}_{e}\nabla\times\rho^{-1}\left(\nabla\times\mathbf{B}\right).
\end{eqnarray}

Equations \eqref{Cont}, \eqref{Mom2} and \eqref{Ohm2} with the total energy,\cite{Kimura14}
\begin{equation}
\label{E}
\mathscr{H}:=\int d^{3}x\left\{\rho\left(\frac{V^{2}}{2}+U\left(\rho\right)\right)+\frac{\mathbf{B}\cdot\mathbf{B}^{\ast}}{2}\right\},
\end{equation}
as Hamiltonian, and the Poisson bracket
\begin{eqnarray} 
\left\{F,G\right\}&=& -\int d^{3}x~\bigg\{\left[F_{\rho} \nabla\cdot G_{\mathbf{V}}+F_{\mathbf{V}}\cdot\nabla G_{\rho}\right]
\label{3dp} \\
& &-\left[\frac{\left(\nabla\times\mathbf{V}\right)}{\rho}\cdot\big(F_{\mathbf{V}}\times G_{\mathbf{V}}\big)\right] 
\nonumber \\ 
& &
-\left[\frac{\mathbf{B}^{\ast}}{\rho}\cdot\big(F_{\mathbf{V}}\times\left(\nabla\times G_{\mathbf{B}^{\ast}}\right)\big)\right] 
\nonumber\\
& & -\left[\frac{\mathbf{B}^{\ast}}{\rho}\cdot\big(\left(\nabla\times F_{\mathbf{B}^{\ast}}\right)\times G_{\mathbf{V}}\big)\right]
\nonumber\\
& &+ d_{i}\left[ \frac{\mathbf{B}^{\ast}}{\rho}\cdot\big(\left(\nabla\times F_{\mathbf{B}^{\ast}}\right)\times\left(\nabla\times G_{\mathbf{B}^{\ast}}\right)\big)\right]
\nonumber\\
& & -d^{2}_{e}\left[ \frac{\left(\nabla\times\mathbf{V}\right)}{\rho}\cdot\big(\left(\nabla\times F_{\mathbf{B}^{\ast}}\right)\times\left(\nabla\times G_{\mathbf{B}^{\ast}}\right)\big)\right]\bigg\}\nonumber
\end{eqnarray}
constitute a noncanonical Hamiltonian system in which the phase space is spanned by the dynamical variables
$\rho, \mathbf{V}$, and $\mathbf{B}^{\ast}$. In \eqref{3dp} $F_{\bm{\xi}}:=\delta F/\delta\bm{\xi}$ denote the functional derivative of the functional $F$ with respect to the dynamical variable $\bm{\xi}$. The  full  Poisson bracket of Eq. \eqref{3dp} and a proof of the Jacobi identity were first given in Ref.~\onlinecite{Abd15}, with further properties and a simplified proof of the Jacobi identity given in Ref.~\onlinecite{Lin15}.  The bracket of Eq. \eqref{3dp} is an extension of the MHD bracket first given in Ref.~\onlinecite{MG80}, amended by the inclusion of two additional terms,  one proportional to $d_i$, accounting  for  the Hall effect,  and  one proportional to $d_e^2$, accounting   for  electron inertia.  

The Poisson bracket \eqref{3dp} has  three independent Casimir invariants, 
\begin{eqnarray}
\label{c1}
\ccal_{1}&=&\int d^{3}x~\mathbf{B}^{\ast}\cdot\left(\mathbf{V}-\frac{d_{i}}{2d^{2}_{e}}\mathbf{A}^{\ast}\right),
\\
\label{c2}
\ccal_{2}&=&\int d^{3}x~ \left[\mathbf{B}^{\ast}\cdot\mathbf{A}^{\ast}+d^{2}_{e}\mathbf{V}\cdot\left(\nabla\times\mathbf{V}\right)\right],
\\
\label{c3}
\ccal_{3}&=&\int d^{3}x~\rho.
\end{eqnarray}
Combining $\ccal_{1}$ and $\ccal_{2}$, produces the ``canonical helicities''
\begin{equation}
\label{c4}
\ccal_{\pm} =\frac{1}{2}\int d^{3}x~\mathbf{P}^{\pm}\cdot\left(\nabla\times \mathbf{P}^{\pm}\right),
\end{equation}
where  $\mathbf{P}^{\pm}=\mathbf{V}+\lambda_{\pm}\mathbf{A}^{\ast},
$ with  
\beq   \label{lambda}
\lambda_{\pm}=\frac{-d_{i}\pm\sqrt{d^{2}_{i}+4d^{2}_{e}}}{2d^{2}_{e}}.
\eeq

%%%%%%%%%%%%%%%%%%%%%%%%%%%%%%%
%%%%%%%%%%%%%%%%%%%%%%%%%%%%%%%
\subsection{Reduced extended MHD }

%%%%%%%%%%%%%%%%%%%%%%%%%%%%%%%
\subsubsection{Direct reduction}

In the incompressible limit, the reduced extended magnetohydrodynamics (RXMHD) can be obtained by writing  $\mathbf{V}$ and $\mathbf{B}$ in the Clebsch-like forms
\begin{eqnarray}
\label{var1}
\mathbf{B}\left(x,y,t\right)&=&\nabla\psi\left(x,y,t\right)\times\widehat{z}+b\left(x,y,t\right)\widehat{z},
\\
\label{var2}
\mathbf{V}\left(x,y,t\right)&=&-\nabla\phi\left(x,y,t\right)\times\widehat{z}+v\left(x,y,t\right)\widehat{z}\,, 
\end{eqnarray}
where $\psi$ and $\phi$ are the flux and stream functions, respectively, and $b$ and $v$ are $\widehat{z}$-components of  these fields. 
From  \eqref{var1},  the current density $\mathbf{J}$ is seen to be given by 
\begin{equation}
\label{Current}
\mathbf{J}=\nabla\times\mathbf{B}=\nabla b\times\widehat{z}-\nabla^{2}\psi\, \widehat{z},
\end{equation}
Upon setting $\rho=1$ and using \eqref{var1} and \eqref{var2},  the $\widehat{z}$-component of Eq. \eqref{Mom} yields
\begin{eqnarray}
\label{R1}
\frac{\partial v}{\partial t}&=&-\left[\phi,v\right]+\left[b,\psi\right]-d^{2}_{e}\left[b,\nabla^{2}\psi\right],
\end{eqnarray}
where $\left[f,g\right]=\nabla f\times\nabla g \cdot \widehat{z}$, 
is the standard canonical Poisson bracket with $x$ and $y$ as canonically conjugate coordinates.
Similarly,  operating with  $\widehat{z}\cdot\curl$ on \eqref{Mom} yields 
\begin{eqnarray}\label{R2}
\frac{\partial \nabla^{2}\phi}{\partial t}&=&-\left[\phi,\nabla^{2}\phi\right]
 -\left[\nabla^{2}\psi,\psi\right]-d^{2}_{e}\left[b,\nabla^{2}b\right]\,,
\end{eqnarray}
and  the $\hat{z}$-component of \eqref{Ohm} is 
\begin{eqnarray}
\label{R3}
-\frac{\partial \psi}{\partial t}+\left[\psi,\phi\right]&=&d_{i}\left[b,\psi\right]-d^{2}_{e}\frac{\partial }{\partial t}\nabla^{2}\psi+d^{2}_{e}\left[\nabla^{2}\psi,\phi\right]\nonumber\\& &+d^{2}_{e}\left[v,b\right]-d_{i}d^{2}_{e}\left[b,\nabla^{2}\psi\right],
\end{eqnarray}
where we made use of the relation $E_{z}=-\partial \psi/\partial t $.
Finally  operating with  $\widehat{z}\cdot\curl$ on \eqref{Ohm} gives
\begin{eqnarray}
\label{R4}
-\frac{\partial b}{\partial t}&&+\left[v,\psi\right]-\left[\phi,b\right]=d_{i}\left[\psi,\nabla^{2}\psi\right]-d^{2}_{e}\frac{\partial }{\partial t}\nabla^{2}b\nonumber\\& &+d^{2}_{e}\left[\nabla^{2}\phi,b\right]+d^{2}_{e}\left[\nabla^{2}b,\phi\right]-d_{i}d^{2}_{e}\left[b,\nabla^{2}b\right].
\end{eqnarray}
Therefore, with the definitions
\begin{eqnarray}
\omega&=&\nabla^{2}\phi\\
  \psi^{\ast}&=&\psi-d^{2}_{e}\nabla^{2}\psi\\   
b^{\ast}&=&b-d^{2}_{e}\nabla^{2}b\,, 
\end{eqnarray}
the  RXMHD equations can be  written as follows:
\begin{eqnarray}
\label{flux}
\frac{\partial \psi^{\ast}}{\partial t}&=&-\left[\phi,\psi^{\ast}\right]-d_{i}\left[b,\psi^{\ast}\right]+d^{2}_{e}\left[b,v\right],
\\
\label{vor}
\frac{\partial \omega}{\partial t}&=&-\left[\phi,\omega\right]-\left[\nabla^{2}\psi,\psi\right]-d^{2}_{e}\left[b,\nabla^{2}b\right],
\\
\label{b}
\frac{\partial b^{\ast}}{\partial t}&=&-\left[\phi,b^{\ast}\right]+d_{i}\left[\nabla^{2}\psi,\psi\right]+\left[v,\psi\right]\nonumber\\& &~~~~~~~~~~+d^{2}_{e}\left[b,\omega\right]+d_{i}d^{2}_{e}\left[b,\nabla^{2}b\right],
\\
\label{v}
\frac{\partial v}{\partial t}&=&-\left[\phi,v\right]+\left[b,\psi^{\ast}\right]\,.
\end{eqnarray}

The Hamiltonian (energy) \eqref{E} in terms of the new variables becomes 
\begin{eqnarray}
\label{RE}
\mathscr{H}:= \frac{1}{2} \int d^2 x\left( - \phi \omega - \lap\psi \pe +b b^* + v^2\right)\,, 
\end{eqnarray}
which can be shown by direct calculation  to be  conserved by the RXMHD system of \eqref{flux}--\eqref{v}.
%%%%%%%%%%%%%%

%%%%%%%%%%%%%%%%%%%%%%%%%%%%%%%
\subsubsection{Reduction via chain rule}

Another way to obtain RXMHD is by Hamiltonian reduction.  With this method  the Poisson bracket \eqref{3dp} is rewritten in terms of the new variables via the functional chain rule (see e.g.\  Refs.~\onlinecite{amp0,amp1} where this is done for MHD).  This method has the advantage of yielding directly the Hamiltonian structure of RXMHD.  

The chain rule proceeds by assuming functionals  obtain their dependence on $\mathbf{V}$ and $\mathbf{B}^{\ast}$ through the new variables $\omega, v, \psi^{\ast}$, and $b^{\ast}$, i.e. 
\begin{equation}
\label{Funct}
F\left[\mathbf{V},\mathbf{B}^{\ast}\right]=\bar{F}\left[\omega, v, \psi^{\ast}, b^{\ast}\right]
\end{equation}
Varying both sides of \eqref{Funct}  gives 
\begin{equation}
\label{VariatV}
\int d^{2}x~F_{\mathbf{V}}\cdot\delta \mathbf{V} 
=\int d^{2}x~\Big(\bar{F}_{\omega}\,  \delta\omega +\bar{F}_{v}\, \delta v \Big),
\end{equation}
while variation of the velocity field $\mathbf{V}$ of \eqref{var2} gives
\begin{equation}
\label{Vvar}
\delta\mathbf{V}=\widehat{z}\times\nabla\delta\phi+\delta v\widehat{z}. 
\end{equation}  
From \eqref{Vvar} we obtain
\begin{equation}
\delta v= \widehat{z}\cdot \delta\mathbf{V} 
\end{equation}
while $\widehat{z} \times \delta\mathbf{V}= -\nabla\delta \phi$.  Thus using $\delta \omega= \nabla \cdot \nabla\delta \phi$, we obtain
\begin{equation}
\label{wvar}
\delta\omega=\nabla\cdot\big(\delta\mathbf{V}\times \widehat{z}\big)\,.
\end{equation}
Upon inserting Eqs. \eqref{Vvar} and \eqref{wvar} into Eq. \eqref{VariatV}, performing an integration by parts,  and using the arbitrariness of $\delta \mathbf{V}$ we obtain
\begin{equation}
\label{FunctV}
 F_{\mathbf{V}}=\nabla \bar{F}_{\omega}\times \hat{z}+\bar{F}_{v}.
\end{equation}
In a similar way we obtain
\begin{equation}
\label{FunctB}
 \nabla\times F_{\mathbf{B}^{\ast}}=\nabla \bar{F}_{b^{\ast}}\times \hat{z}+\bar{F}_{\psi^{\ast}}\hat{z}.
\end{equation}

Now we are in position to use  \eqref{FunctV} and \eqref{FunctB}  to reduce the Poisson bracket of  \eqref{3dp} to one in terms of the reduced variables.  This calculation gives 
\begin{eqnarray}    \label{pb2d}
\{F,G\}&=&\int d^2 x \bigg\{ \psi^{\ast} \bigg(\left[F_{\omega} , G_{\pe}\right]+ \left[F_{\pe} , G_{\omega}\right]
\nonumber\\
&+&\left[F_v , G_{b^*}\right]+\left[F_{b^*} , G_v\right]
  \\
&-&d_i \left(\left[F_{\pe} , G_{b^*}\right] +\left[F_{b^*} , G_{\pe}\right]\right)\bigg)
 \nonumber\\
 &+&\omega\Big( \left[F_{\omega} , G_{\omega}\right] + \de \left[F_{b^*} , G_{b^*}\right]\Big)
 \nonumber\\
 &+& b^*\Big(\left[F_{\omega} , G_{b^*}\right]+\left[F_{b^*} , G_{\omega}\right] - d_i \left[F_{b^*} , G_{b^*}\right]\Big) 
\nonumber\\
&+& v\Big(\left[F_{\omega} , G_v\right] + \left[F_v , G_{\omega}\right]
\nonumber\\
&+& \de \big(\left[F_{\pe} , G_{b^*}\right] + \left[F_{b^*} , G_{\pe}\right]\big)\Big)\bigg\},
\nonumber
\end{eqnarray}
where, consistent with the representation \eqref{var2}, we have removed the $\rho$ dependence and used  the relation
\begin{equation}
\label{fgh}
\int d^{2}x~f\left[g,h\right]=\int d^{2}x~h\left[f,g\right]=\int d^{2}x~g\left[h,f\right],
\end{equation}
valid for  generic functions $f$, $g$ and $h$ and appropriate boundary conditions.  Here and henceforth,   we drop the bars on the functionals. 

The above bracket \eqref{pb2d} with the Hamiltonian \eqref{RE} produces the equations of motion \eqref{flux}--\eqref{v} in the form  
$\partial \zeta/\partial t=\left\{\zeta,\mathscr{H}\right\}$, where $\zeta=\left(\psi^{\ast},\omega,b^{\ast},v\right)^{t}$ denotes the dynamical variables of the system.
 
%%%%%%%%%%%%%%%%%%%%%%%%%%%%%%%
%%%%%%%%%%%%%%%%%%%%%%%%%%%%%%%
\subsection{Jacobi identity}\label{Jacobi}

As a further check that the set of equations \eqref{flux}--\eqref{v} with the Hamiltonian \eqref{RE}  constitutes a noncanonical Hamiltonian system with  Poisson bracket \eqref{pb2d}, we verify the following requisite bracket properties: 
antisymmetric
\begin{itemize}
\item antisymmetry
\[
\left\{F,G\right\}=-\left\{G,F\right\},
\]
\item Leibniz property
\[
\left\{FG,H\right\}=F\left\{G,H\right\}+G\left\{F,H\right\},
\]
\item Jacobi identity
\[
\left\{F,\left\{G,H\right\}\right\}+\left\{H,\left\{F,G\right\}\right\}+\left\{G,\left\{H,F\right\}\right\}=0,
\]
\end{itemize}
Assuming  boundary conditions such that surface terms vanishes, as would be the case for periodic boundary conditions, we can easily demonstrate the first two properties. However, the proof of Jacobi identity is more difficult.  A direct proof is tedious, but instead we can follow  the general theory of Ref.~\onlinecite{Thi00}.  Using  $\zeta=\left(\psi^{\ast},\omega,b^{\ast},v\right)^{t}$ with each field being indexed by $\zeta^{\mu}$,  $\mu=1, \cdots ,4$,  we can write \eqref{pb2d} in the form
\begin{equation}
\{F,G\}=\int d^2 x [F_{\mu}, F_\nu] W^{\mu \nu}_{\gamma} \zeta^\gamma,
\end{equation}
where $F_{\mu}=\delta F/\delta \zeta^\mu$ and the  quantities $W^{\mu \nu}_\gamma$ are symmetric  in their upper indices.  Considering the  $W^{\mu \nu}_\gamma$  as a family of matrices indexed by $\nu$,   the Jacobi identity is satisfied  if and only if  the following matrices  pairwise commute:
\begin{eqnarray}
W^{(\omega)}&=&\begin{pmatrix} 1 & 0 & 0 & 0 & \\  0 & 1 & 0 & 0 & \\ 0 & 0& 1 & 0& \\ 0 & 0& 0& 1 & \end{pmatrix}, 
W^{(\pe)}=  \begin{pmatrix} 0 & 0 & 0 & 0 & \\  1 & 0 & -d_i & 0 & \\ 0 & 0& 0 & 0& \\ 0 & 0& \de& 0 & \end{pmatrix},  
\nonumber \\
W^{(b^*)}&=&\begin{pmatrix} 0 & 0 & \de & 0 & \\  0 & -d_i & 0 & 1 & \\ 1 & 0& -d_i & 0& \\ 0 & \de & 0&0 & \end{pmatrix}, 
W^{(v)}=\begin{pmatrix} 0 & 0 & 0 & 0 & \\  0 & 0 & 1 & 0 & \\ 0 & 0& 0 & 0& \\ 1 & 0& 0& 0 & \end{pmatrix},\nonumber
\end{eqnarray}
which follows from a relatively easy calculation. Consequently, the Poisson bracket \eqref{pb2d} satisfies the Jacobi identity.

%%%%%%%%%%%%%%%%%%%%%%%%%%%%%%%
%%%%%%%%%%%%%%%%%%%%%%%%%%%%%%%
\subsection{Remarkable transformations}

In Ref.~\onlinecite{Lin15} it was shown that the Poisson bracket \eqref{3dp}  follows from a remarkable sequence of variable and parameter transformations of a basic bracket for  Hall MHD.  This led a dramatically simplified calculation for the Jacobi identity and quite naturally to the Casimir invariants. We will show that the reduced Poisson bracket of    \eqref{pb2d}  possesses analogous  transformations. 

Specifically, the bracket (\ref{3dp}) maps into the  Poisson bracket of Hall MHD in terms of the field $\bbl$, when one carries out the transformation
\ben
\bbl=\mathbf{B}^*+\lambda_{\pm}^{-1} \curl  \bv,
\een
The analogous transformation in  our 2D case would then  be of the form,
\ben
\bbl=\nabla  \pl  \times \hz +  \bl \hz
\een
and this suggests the  change of variables  
\begin{equation}
\pl=\pe+v/\lambda_{\pm}, \qquad \bl=b^*+\omega/\lambda_{\pm}.
\end{equation}
With this change of variables,  the bracket (\ref{pb2d}) becomes
\begin{eqnarray}
\{F,G\}&=&\int\! \!d^2 x \Bigg\{\psi_{\lambda_{\pm}}\!\bigg(\!\!\left(\frac{2}{\lambda_{\pm}}-d_{i}\right)\!\!\Big(\!\left[F_{\psi_{\lambda_{\pm}}},	G_{b_{\lambda_{\pm}}}\right]
\label{bhall}\\
&+&\left[F_{b_{\lambda_{\pm}}},	G_{\psi_{\lambda_{\pm}}}\right]\Big)
+ \left[F_{\psi_{\lambda_{\pm}}},	G_{\omega}\right]+ \left[F_{\omega},	G_{\psi_{\lambda_{\pm}}}\right]
\nonumber\\
&+&\left[F_{v},	G_{b_{\lambda_{\pm}}}\right] +\left[F_{b_{\lambda_{\pm}}},	G_{v}\right]\bigg)
+ \omega\left[F_{\omega}G_{\omega}\right]
\nonumber\\
&+&v\Big(\left[F_{\omega},	G_{v}\right]+\left[F_{v},	G_{\omega}\right]\Big)
+b_{\lambda_{\pm}}\bigg(\left[F_{\omega},	G_{b_{\lambda_{\pm}}}\right] 
\nonumber\\
&+& \left[F_{b_{\lambda_{\pm}}},	G_{\omega}\right]+\left(\frac{2}{\lambda_{\pm}}-d_{i}\right)\left[F_{b_{\lambda_{\pm}}},	G_{b_{\lambda_{\pm}}}\right]\bigg)\Bigg\}\,.
\nonumber
\end{eqnarray}
Note that one obtains the bracket (\ref{bhall}) for either choice of the values of $\lambda_{\pm}$ in Eq. (\ref{lambda}).
Also, note that the bracket (\ref{bhall}) is identical to the Poisson bracket identified by Eqs.~(43)--(44) in Ref.~\onlinecite{Haz87} if one replaces, in the latter bracket, $\beta$ with $-1$ and 
$2\delta \beta$ with $2/\lambda_{\pm} - d_i$. We have thus shown that the bracket (\ref{pb2d}) can be transformed, by means of an invertible change of variables, into a known Poisson bracket for which the Jacobi identity has already been proven.  Consequently, this serves as an  alternative verification  that  the bracket (\ref{pb2d}) satisfies the Jacobi identity. We remark that the model in Ref.~\onlinecite{Haz87} (in the 2D cold ion limit with no magnetic
curvature), is isomorphic to 2D incompressible Hall MHD, which is consistent with the above mentioned general result of Ref.~\onlinecite{Lin15}.

Given the relationship to the results of Ref.~\onlinecite{Haz87} we can immediately  identify the Casimir invariants and normal fields, a special class of field variables, which we consider next.

%%%%%%%%%%%%%%%%%%%%%%%%%%%%%%%
%%%%%%%%%%%%%%%%%%%%%%%%%%%%%%%
%%%%%%%%%%%%%%%%%%%%%%%%%%%%%%%
\section{Normal fields and Casimir invariants}
\label{sec:normalCas}

%%%%%%%%%%%%%%%%%%%%%%%%%%%%%%%
%%%%%%%%%%%%%%%%%%%%%%%%%%%%%%%
\subsection{Normal fields}
\label{ssec:normalF}

The four-field bracket of \eqref{bhall} is complicated, as one might expect  considering the physics described by the  RXMHD model.  However,  as described in  Ref.~\onlinecite{Thi00}, noncanonical brackets can be mapped by coordinate changes  into special simplified forms.  For systems of four fields, there are only a few such simplified forms.  The fields in which the bracket is simplified are called normal fields -- for  the present case they are given by
\begin{align}
\begin{split}
\psi_+=&\pl , \qquad \psi_-=\pl -\left(\frac{2}{\lambda_+} - d_i\right) v, \\
b_+=&\bl , \qquad b_-=\bl- \left(\frac{2}{\lambda_+} - d_i\right)\omega.
\end{split}
\end{align}
In terms of these normal fields  the bracket (\ref{bhall}) becomes
\begin{eqnarray}
 \label{pnormal}
\{F,G\}&=&\left(\frac{2}{\lambda_+} - d_i\right)\!\int \!\!d^2 x \bigg\{ 
\psi_{+} \Big( [F_{\psi_+} , G_{b_+}] \nonumber\\
&+& [F_{b_+} , G_{\psi_+}]\Big) +b_{+} [ F_{b_{+}} , G_{b_{+}}] -b_{-}[F_{b_{-}} , G_{b_{-}}]
\nonumber\\
&-& \psi_{-}\Big( [F_{\psi_{-}} , G_{b_{-}}]+[F_{b_{-}} , G_{\psi_{-}}]\Big) \bigg\}\,, 
\end{eqnarray}
a form that is  the  direct sum of two semidirect product brackets (see Ref.~\onlinecite{Thi00}).  In terms of the normal fields $\psi_{\pm} , b_{\pm}$  the corresponding Casimirs for this bracket are known to be 
\begin{equation}
\label{c3456}
C_{1,2}=\int d^2 x ~ \mathcal{F}_{\pm}\left(\psi_{\pm}\right),\\~~~~~~~
C_{3,4}=\int d^2 x ~ b_{\pm}\mathcal{G}_{\pm}\left(\psi_{\pm}\right),
\end{equation}
with $\mathcal{F}_{\pm}$ and $\mathcal{G}_{\pm}$ arbitrary functions. 

We remark that, in the 2D incompressible limit, the Casimir invariants $\ccal_{1,2}$ of XMHD reduce to 
\begin{align}
&\int d^{2}x~\mathbf{B}^{\ast}\cdot\left(\mathbf{V}-\frac{d_{i}}{2d^{2}_{e}}\mathbf{A}^{\ast}\right)
\nonumber\\
&=\int d^{2}x\left(\omega\psi^{\ast}+vb^{\ast}-\frac{d_{i}}{2d^{2}_{e}}\psi^{\ast}b^{\ast}\right),
\end{align}
and
\begin{align}
&\int d^{2}x~\big[\mathbf{B}^{\ast}\cdot\mathbf{A}^{\ast}+d^{2}_{e}\textbf{V}\cdot\left(\nabla\times\mathbf{V}\right)\big]
\nonumber\\
&= \int d^{2}x\left(\psi^{\ast}b^{\ast}+d^{2}_{e}\, v \omega\right),
\label{c2d2}
\end{align}
respectively. Such Casimir invariants indeed correspond to linear combinations of the Casimir invariants $C_{3,4}$ of Eq. (\ref{c3456}), for the particular choice $\mathcal{G}_{\pm}=\psi_{\pm}$.
This shows how the Casimir invariants of XMHD are related to those of RXMHD.

 We remark that, in Ref.~\onlinecite{And14}, a system isomorphic to RXMHD was studied but only three out of the infinite number of  invariants of the model were presented.

Noting that $2/ \lambda_+ -d_i = 1/ \lambda_+ -1/ \lambda_-$, we find  that the normal fields are related to  the original variables by
\beq
\label{peb}
\psi_{\pm} = \pe +\frac{v}{\lambda_{\pm}} \qquad \mathrm{and}\qquad
 b_{\pm}= b^*+\frac{\omega}{\lambda_{\pm}}.
\eeq
In terms of the normal fields the RXMHD system obtains the perspicuous form
\ben
\frac{\pa \psi_{\pm}}{\pa t}+\left[\phi_{\pm} , \psi_{\pm}\right]&=&0,  \label{psipm} \\
\frac{\pa b_{\pm}}{\pa t} +\left[\phi_{\pm} , b_{\pm}\right]&=& \lambda_{\pm} [ \psi_{\pm} , \psi], \label{bpm}
\een
where $\phi_{\pm}=\phi-d_e^2 \lambda_{\pm} b$.  Here we also made use of the relation $d_i- 1/ \lambda_{\pm}=- d_e^2 \lambda_{\pm}$.  From Eqs.~(\ref{psipm}) and (\ref{bpm}) it  emerges that $\psi_{\pm}$ are Lagrangian invariants of the model, reminiscent of Ohm's law for reduced MHD (RMHD), whereas the equations describing $b_{\pm}$ are reminiscent of the RMHD vorticity equation. 

We remark that, by expanding in the limit $d_e^2 / d_i^2 \rightarrow 0$, one can obtain the following relations:
\ben
\psi_+ \simeq \psi + d_i v_{iz}, \qquad   \psi_- \simeq \psi -\frac{d_e^2}{d_i} v_{ez}, \label{n1} \\
\hz\times \nabla \phi_+ \simeq \mathbf{v}_{i \perp}, \qquad \hz\times \nabla \phi_- \simeq \mathbf{v}_{e \perp}, \label{n2} \\
b_+ \simeq b + d_i \omega_i, \qquad b_- \simeq b - \frac{d_e^2}{d_i} \omega_e, \label{n3} \\
\phi \simeq \phi_+ + \frac{d_e^2}{d_i^2} \phi_- , \qquad b \simeq \frac{\phi_- - \phi_+}{d_i}, \label{n4}
\een
where $v_{iz}$ and $v_{ez}$ are the $z$-components of the ion and electron fluid velocities, (so that $\nabla^2 \psi=(v_{ez} - v_{iz})/d_i$ and $v \simeq v_{iz}+(d_e^2 / d_i^2) v_{ez})$, $\mathbf{v}_{i \perp}$ and $\mathbf{v}_{e \perp}$ are the ion and electron perpendicular fluid velocities, whereas $\omega_{i,e}=\hz\cdot \nabla \times \mathbf{v}_{i,e \perp}$ are the $z$ components of the corresponding vorticities. From Eqs.~(\ref{n1})--(\ref{n3}) it emerges then
that $\psi_{\pm}$ correspond to the $z$-components of the canonical momenta for ions and electrons. These are advected, according to Eqs.~(\ref{psipm}), by the perpendicular ion and electron velocities, respectively. The normal fields $b_{\pm}$, on the other hand, represent some generalized vorticities, analogous to the generalized vorticity of Hall-MHD.   

The Hamiltonian (\ref{RE}) can be expressed in terms of the normal fields by making use of the following transformations:
\ben
\psi^{\ast}=\frac{\lambda_+ \psi_+ - \lambda_- \psi_-}{\lambda_+ - \lambda_-}, \qquad v=\frac{\psi_+ - \psi_-}{d_e^2 (\lambda_+ - \lambda_-)}, \\
b^*=\frac{\lambda_+ b_+ - \lambda_- b_-}{\lambda_+ - \lambda_-}, \qquad \omega=\frac{b_+ - b-}{d_e^2 (\lambda_+ - \lambda_-)},
\een
and introducing the linear operator $\mcal$, such that $\psi^{\ast}= \mcal \psi$ and $b^*= \mcal b$. Assuming this operator is invertible,
one can then write 
\beq
\psi= \mcal^{-1}  \frac{\lambda_+ \psi_+ - \lambda_- \psi_-}{\lambda_+ - \lambda_-}, \quad b=\mcal^{-1} \frac{\lambda_+ b_+ - \lambda_- b_-}{\lambda_+ - \lambda_-},
\eeq
and replace these expressions in (\ref{RE}). The resulting functional is
\beq  \label{normalH}
\begin{split}
 \mathscr{H}=\frac{1}{2}\int d^2 x &\bigg( -\frac{b_{+} - b{-}}{d_e^2 \left(\lambda_{+} - \lambda_{-}\right)}\nabla^{-2}\frac{b_{+} - b{-}}{d_e^2 \left(\lambda_{+} - \lambda_{-}\right)}\\
 &+\frac{\lambda_{+} b_{+} - \lambda_{-} b_{-}}{\lambda_{+} - \lambda_{-}}\mcal^{-1}\frac{\lambda_{+} b_{+} - \lambda_{-} b_{-}}{\lambda_{+} - \lambda_{-}} \\
 &-\frac{\lambda_{+} \psi_{+} - \lambda_{-} \psi_{-}}{\lambda_{+} - \lambda_{-}} \nabla^2 \mcal^{-1}\frac{\lambda_{+} \psi_{+} - \lambda_{-} \psi_{-}}{\lambda_{+} - \lambda_{-}}\\
 &+\frac{\psi_{+}^{2} - 2 \psi_{+}\psi_{-} +\psi_{-}^{2}}{d_e^4 \left(\lambda_{+} - \lambda_{-}\right)^2}\bigg).
\end{split}
\eeq

Because the Hamiltonian of \eqref{normalH} is complicated it may be  more straightforward to consider that of \eqref{RE} in terms of the original variables.  An approximate form can be obtained by neglecting again $d_e^2/d_i^2$ when compared to terms of order unity, and making use of the relations (\ref{n2}), (\ref{n4}), and the relation $\nabla^2 \psi=(v_{ez} - v_{iz})/d_i$.  This leads to  the following approximate expression for the Hamiltonian:
\begin{eqnarray}
 \label{happr}
\mathscr{H}&\simeq&\frac{1}{2}\int d^2 x \bigg( \vert \nabla \psi \vert^2 +b^2 + v_{i \perp}^2 + v_{iz}^2 
\nonumber\\
&&\hspace{ 2cm} + \frac{d_e^2}{d_i^2}\left(v_{e \perp}^2 +v_{ez}^2\right)\bigg).
\end{eqnarray}
The expression (\ref{happr}) shows that the Hamiltonian  is nearly given by the sum of magnetic energy (the first two terms on the right-hand side of (\ref{happr})),  with the ion kinetic energy (third and fourth terms) and  the electron kinetic energy (fifth and sixth terms).

%%%%%%%%%%%%%%%%%%%%%%%%%%%%%%%
%%%%%%%%%%%%%%%%%%%%%%%%%%%%%%%

%%%%%%%%%%%%%%%%%%%%%%%%%%%%%%% 
%%%%%%%%%%%%%%%%%%%%%%%%%%%%%%% 
%%%%%%%%%%%%%%%%%%%%%%%%%%%%%%% 
\section{Limits of RXMHD}
\label{sec:limits}

%%%%%%%%%%%%%%%%%%%%%%%%%%%%%%%
%%%%%%%%%%%%%%%%%%%%%%%%%%%%%%%
\subsection{2D incompressible Hall MHD}

If we set $d_e=0$ in Eqs.~\eqref{flux}--\eqref{v}, we obtain the 2D incompressible Hall MHD system
\ben
\frac{\pa \psi}{\pa t}&=&-[\phi , \psi]- d_i [b,\psi],  \label{psih}\\
\frac{\pa \omega}{\pa t}&=&-[\phi , \omega]-[\lap \psi , \psi],\\
\frac{\pa b}{\pa t}&=&-[\phi , b]+ d_i [\lap \psi ,\psi] + [v,\psi] ,\\
\frac{\pa v}{\pa t}&=&-[\phi ,v]+[b, \psi]   \label{vh}.
\een
As  anticipated above, this model  is also Hamiltonian, with Hamiltonian functional 
\begin{equation}
H=\frac{1}{2} \int d^2 x\left( \vert \nabla \phi \vert^2 + \vert \nabla \psi \vert^2 + b^2 +v^2 \right),
\end{equation}
and Poisson bracket
\begin{equation}  \label{2dHall}
\begin{split}
\{F,G\}=&\int d^2 x \bigg\{ \psi \bigg(\left[F_{\omega} , G_{\psi}\right]+ \left[F_{\psi} , G_{\omega}\right]+\left[F_v , G_{b}\right]\\&+\left[F_{b} , G_v\right]-d_i \left(\left[F_{\psi} , G_{b}\right] +\left[F_{b} , G_{\pe}\right]\right)\bigg)
 \\&+\omega \left[F_{\omega} , G_{\omega}\right] + v\Big(\left[F_{\omega} , G_v\right] + \left[F_v , G_{\omega}\right]\Big)\\& +b\Big(\left[F_{\omega} , G_{b}\right]+\left[F_{b} , G_{\omega}\right] - d_i \left[F_{b} , G_{b}\right]\Big)\bigg\}.
\end{split}
\end{equation}
This system, which  previously appeared in Ref.~\onlinecite{Haz87}, has the Casimirs
\begin{eqnarray}
C_{1}&=& \int d^{2}x~\mathcal{K}\left(\psi\right),
\\
C_{2}&=& \int d^{2}x~b~\mathcal{S}\left(\psi\right),
\\
C_{3}&=& \int d^{2}x~\mathcal{T}\left(\psi_{H}\right),
\\
C_{4}&=& \int d^{2}x~b_{H}~\mathcal{R}\left(\psi_{H}\right),
\end{eqnarray}
where $b_{H}=b+d_i \omega$, $\psi_{H}=\psi + d_i v$ and $\mathcal{K}, \mathcal{S}, \mathcal{T} $ and $\mathcal{R}$ are arbitrary functions. In particular, for $\mathcal{S}=\psi$ and $\mathcal{R}=\psi_H$ one 
retrieves the 2D incompressible versions of the functionals
\begin{equation}
\int d^{3}x~\mathbf{A}\cdot\mathbf{B},
\end{equation}
and
\begin{equation}
\int d^{2}x\left(\mathbf{A}+d_{i}\mathbf{V}\right)\cdot\left(\mathbf{B}+d_{i}\nabla\times\mathbf{V}\right),
 \end{equation}
 respectively, which are Casimir invariants for 3D Hall MHD, corresponding to magnetic helicity and to a generalized magnetic helicity.
%%%%%%%%%%%%%%%%%%%%%%%%%%%%%%%
%%%%%%%%%%%%%%%%%%%%%%%%%%%%%%%
\subsection{2D incompressible inertial MHD}

If we set $d_i=0$ in Eqs.~\eqref{flux}--\eqref{v} while retaining $d_e$,  we obtain  the 2D incompressible inertial MHD system 
\begin{eqnarray}
\label{Iflux}
\frac{\partial \psi^{\ast}}{\partial t}&=&-\left[\phi,\psi^{\ast}\right]+d^{2}_{e}\left[b,v\right],
\\
\label{Ivor}
\frac{\partial \omega}{\partial t}&=&-\left[\phi,\omega\right]-\left[\nabla^{2}\psi,\psi\right]-d^{2}_{e}\left[b,\nabla^{2}b\right],
\\
\label{Ib}
\frac{\partial b^{\ast}}{\partial t}&=&-\left[\phi,b^{\ast}\right]+\left[v,\psi\right]+d^{2}_{e}\left[b,\omega\right],
\\
\label{Iv}
\frac{\partial v}{\partial t}&=&-\left[\phi,v\right]+\left[b,\psi^{\ast}\right].
\end{eqnarray}
In this limit the Hamiltonian \eqref{RE} does not change, but the Poisson brackets becomes
\begin{equation}  \label{2dI}
\begin{split}
\{F,G\}=\int d^2 x &\bigg\{ \psi^{\ast} \bigg(\left[F_{\omega} , G_{\pe}\right]+ \left[F_{\pe} , G_{\omega}\right]\\&+\left[F_v , G_{b^*}\right]+\left[F_{b^*} , G_v\right]\bigg)
 \\&+\omega\Big( \left[F_{\omega} , G_{\omega}\right] + \de \left[F_{b^*} , G_{b^*}\right]\Big)\\& +b^*\Big(\left[F_{\omega} , G_{b^*}\right]+\left[F_{b^*} , G_{\omega}\right] \Big) \\& + v\Big(\left[F_{\omega} , G_v\right] + \left[F_v , G_{\omega}\right]\\&+ \de \big(\left[F_{\pe} , G_{b^*}\right] + \left[F_{b^*} , G_{\pe}\right]\big)\Big)\bigg\}.
\end{split}
\end{equation}
We can easily proof that the above system is Hamiltonian through one of the methods discussed in Sec. \ref{Jacobi}.

It may seem odd to retain $d_e$ while dropping $d_i$, since they scale with the mass ration, but this limit may make sense in a different ordering.\cite{Kimura14}

The Poisson bracket (\ref{2dI}) possesses the following four families of Casimir invariants:
\begin{eqnarray}
C_{1,2}&=&\int d^{2}x~\mathcal{Y}_{\pm}\left(\psi^{i}_{\pm}\right),\\
C_{3,4}&=&\int d^{2}x~b^{i}_{\pm}\mathcal{P}_{\pm}\left(\psi^{i}_{\pm}\right),  \label{c34}
\end{eqnarray}
where $b^{i}_{\pm}=b^{\ast}\pm d_{e} \omega, \psi^{i}_{\pm}=\psi^{\ast}\pm d_{e} v$ and $\mathcal{Y}_\pm$ and $\mathcal{P}_\pm$ are arbitrary functions.

3D inertial MHD, which  previously appeared in Ref.~\onlinecite{Lin15}, has the Casimirs
\begin{equation}
\label{c2d1}
\int d^{3}x~\mathbf{B}^{\ast}\cdot \mathbf{V}
\end{equation}
and
\begin{equation}
\int d^{3}x~\big[\mathbf{B}^{\ast}\cdot\mathbf{A}^{\ast}+d^{2}_{e}\textbf{V}\cdot\left(\nabla\times\mathbf{V}\right)\big],
\end{equation}
which, in their 2D incompressible limit, become
\begin{equation}
\int d^2 x \, (\omega \psi^* + b^* v),
\end{equation}
and
\begin{equation}
\int d^2 x \, (\psi^* b^* + d_e^2 v ~\omega ),
\end{equation}
respectively.
These are linear combinations of 
\begin{equation}
\int d^2 x \, b_{\pm}^i \psi_{\pm}^i,
\end{equation}
corresponding to the Casimir invariants $C_{3,4}$ of Eq. (\ref{c34}) for the choice $\mathcal{P}_{\pm}=\psi_{\pm}^i$.

%%%%%%%%%%%%%%%%%%%%%%%%%%%%%%%
%%%%%%%%%%%%%%%%%%%%%%%%%%%%%%%
\subsection{2D incompressible ideal MHD}

The 2D incompressible  ideal MHD system can be obtained by setting $d_i=d_e=0$ in Eqs.~\eqref{flux}--\eqref{v}, giving 
\begin{eqnarray}
\label{idealflux}
\frac{\partial \psi}{\partial t}&=&-\left[\phi,\psi\right],
\\
\label{idealvor}
\frac{\partial \omega}{\partial t}&=&-\left[\phi,\omega\right]-\left[\nabla^{2}\psi,\psi\right],
\\
\label{idealb}
\frac{\partial b}{\partial t}&=&-\left[\phi,b\right]+\left[v,\psi\right],
\\
\label{idealv}
\frac{\partial v}{\partial t}&=&-\left[\phi,v\right]+\left[b,\psi\right].
\end{eqnarray}
Reduced ideal MHD has energy
\begin{equation}
H=\frac{1}{2} \int d^2 x\left( \vert \nabla \phi \vert^2 + \vert \nabla \psi \vert^2 + b^2 +v^2 \right)\,,
\end{equation}
and the  Poisson bracket
\begin{equation}  \label{2dideal}
\begin{split}
\{F,G\}=\int d^2 x &\Big\{ \psi \Big(\left[F_{\omega} , G_{\psi}\right]+ \left[F_{\psi} , G_{\omega}\right]+\left[F_v , G_{b}\right]\\&+\left[F_{b} , G_v\right]\Big)
+\omega \left[F_{\omega} , G_{\omega}\right]  \\&+ v\Big(\left[F_{\omega} , G_v\right] + \left[F_v , G_{\omega}\right]\Big)\\& +b\Big(\left[F_{\omega} , G_{b}\right]+\left[F_{b} , G_{\omega}\right]\Big)\Big\}.
\end{split}
\end{equation}

This reduced ideal MHD model, which  previously appeared in Ref.~\onlinecite{Haz87}, has the Casimirs
\begin{eqnarray}
C_{1}&=& \int d^{2}x~\mathcal{I}\left(\psi\right),
\\
C_{2}&=& \int d^{2}x~b~\mathcal{O}\left(\psi\right),  \label{cas2}
\\
C_3 &=& \int d^2 x ~v ~ \mathcal{Q}(\psi),
\\
C_4 &=& \int d^2 x ~ (\omega ~\mathcal{U} (\psi) + b~v ~\mathcal{U}' (\psi)),  \label{cas4}
\end{eqnarray}
where $\mathcal{I}, \mathcal{O}, \mathcal{Q}, \mathcal{U}$ are arbitrary functions and the prime symbol denotes derivative with respect to the argument of the function.

With the choices $\mathcal{O}=\psi$ and $\mathcal{U}=\psi$ one retrieves, from $C_2$ and $C_4$ in Eqs. (\ref{cas2}) and (\ref{cas4}), the 2D incompressible versions of the magnetic helicity
\begin{equation}
\int d^{3}x~\mathbf{A}\cdot\mathbf{B}
\end{equation}
and of the cross-helicity
\begin{equation}
\int d^{3}x~\mathbf{V}\cdot\mathbf{B}
\end{equation}
of 3D ideal MHD.

%%%%%%%%%%%%%%%%%%%%%%%%%%%%%%%
%%%%%%%%%%%%%%%%%%%%%%%%%%%%%%%
%%%%%%%%%%%%%%%%%%%%%%%%%%%%%%%
\section{Numerical results}
\label{sec:numerics}

%%%%%%%%%%%%%%%%%%%%%%%%%%%%%%%
%%%%%%%%%%%%%%%%%%%%%%%%%%%%%%%
\subsection{Linear analysis}

Before describing our nonlinear simulations, we perform  a simple linear stability analysis in Sec.~\ref{sssec:basic} to verify that RXMHD contains the basic whistler and cyclotron waves.   This is followed by an investigation of collisionless tearing modes in our nonlinear simulation geometry, which serves as an introduction to our nonlinear numerical results. 
 
%%%%%%%%%%%%%%%%%%%%%%%%%%%%%%%
\subsubsection{Basic modes}
\label{sssec:basic}

We linearize the  RXMHD  equations of \eqref{vor}--\eqref{b}  to investigate the basic modes it contains. Upon assuming a magnetostatic equilibrium state corresponding to a unit vector $\hat{b}_0$  in $x-y$ plane, we expand all quantities as $  \psi=\tilde\psi\exp\left(i\mathbf{k}_{\bot}.\mathbf{r}_{\bot}-i w t\right)$, where $w$ is the angular frequency and $\mathbf{k}_{\bot}$ is the perpendicular  wavenumber, to obtain 
\begin{eqnarray}
\label{l1}
&&  \tilde \phi=\frac{\left(\hat{b}_0\cdot \mathbf{k}_{\bot}\right)}{w} \tilde \psi
\qquad 
 \tilde v=-\frac{\left(\hat{b}_0\cdot \mathbf{k}_{\bot}\right)}{w}  \tilde b\nonumber\\
&& w\left(1+d^{2}_{e}k^{2}_{\bot}\right) \tilde \psi=\left(\hat{b}_0\cdot \mathbf{k}_{\bot}\right)  \tilde \phi+d_{i}\left(\hat{b}_0\cdot \mathbf{k}_{\bot}\right) \tilde  b
\nonumber\\
&&w\left(1+d^{2}_{e}k^{2}_{\bot}\right)  \tilde b=\left(\hat{b}_0\cdot \mathbf{k}_{\bot}\right)  \tilde v -d_{i}k^{2}_{\bot}\left(\hat{b}_0\cdot \mathbf{k}_{\bot}\right)  \tilde \psi\,.
\nonumber
\end{eqnarray}
Manipulation of the above yields
\begin{eqnarray}
\label{l5}
w\left(1+d^{2}_{e}k^{2}_{\bot}\right)\tilde\psi&=&\frac{\left(\hat{b}_0\cdot \mathbf{k}_{\bot}\right)^{2}}{w} \tilde  \psi+d_{i}\left(\hat{b}_0\cdot \mathbf{k}_{\bot}\right) \tilde  b,
\nonumber\\
\label{l6}
w\left(1+d^{2}_{e}k^{2}_{\bot}\right) \tilde  b&=&\frac{\left(\hat{b}_0\cdot \mathbf{k}_{\bot}\right)^{2}}{w} \tilde  b+d_{i}k^{2}_{\bot}\left(\hat{b}_0\cdot \mathbf{k}_{\bot}\right) \tilde  \psi\,,
\nonumber
\end{eqnarray}
whence we obtain the dispersion relation of RXMHD  
\begin{equation}
\left\{w^{2}\left(1+d^{2}_{e}k^{2}_{\bot}\right)-k^{2}_{\bot}\cos^{2}\theta \right\}^{2}=w^{2}d^{2}_{i}k^{4}_{\bot}\cos^{2}\theta \,, \nonumber
\end{equation}
where $\theta$ is the angle between $\hat{b}_0$ and $\mathbf{k}_{\bot}$.

\begin{figure}[h!]
\centering
	  \includegraphics[width=0.47\textwidth]{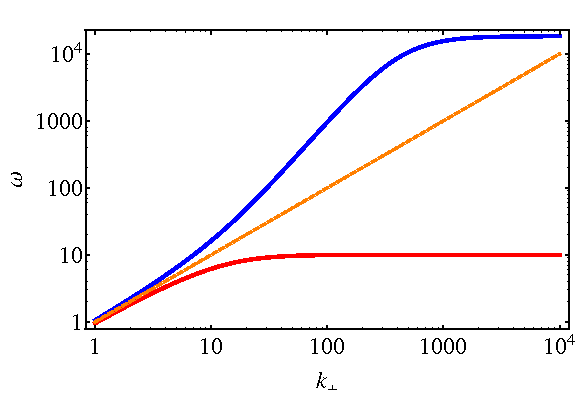}
\caption{The dispersion relation $\left(w \right)$ profiles for $\theta=0$, $d_{i}=0.1$ and $d_{e}=0.0233$.  The upper branch corresponds to whistler waves, while the lower branch represents ion cyclotron waves.  The dashed  reference line corresponding to  ideal Alfv\'en waves  which in dimensionless units is $w=k_{\bot}$. }
 \label{Figg.1}
\end{figure}

As expected, this linear dispersion relation is coincident with the 3D nonlinear dispersion relation of XMHD.\cite{Hamdi16}  In Fig.~\ref{Figg.1}, the upper branch represents whistler waves, whilst the lower branch represents ion cyclotron waves. We can also observe that both branches saturate,  at  the electron gyrofrequency and ion gyrofrequency,  respectively. 
 
%%%%%%%%%%%%%%%%%%%%%%%%%%%%%%%
\subsubsection{Collisionless tearing modes} 
 \label{sec:rlin}

The XMHD model of (\ref{flux})--(\ref{v}) can describe various instabilities, including  collisionless tearing modes induced by the presence of electron inertia, which breaks the usual MHD frozen-in condition as evidenced by  Eqs.~(\ref{flux})
and (\ref{b}).

In order to investigate collisionless tearing, we suppose  a resonant surface is  located at $x=0$ 
 and we choose an equilibrium around $x=0$  given by 
\beq  \label{eqlin}
\psi_{eq}=-x^2, \qquad b_{eq}=0, \qquad \phi_{eq}=0, \qquad v_{eq}=0. 
\eeq
Then upon linearizing the  system of  (\ref{flux})--(\ref{v}) about this equilibrium,  assuming   solutions of the form
$\psi(x,y,t)=\tilde{\psi}(x)\mathrm{e}^{i ky+\gamma t}$ with analogous expressions for $\phi, v,$ and $b$,
where the constants $\gamma$ and $k$ indicate the growth rate and the wave number of the perturbation, respectively, 
we obtain
\ben
g((1+k^2 d_e^2) \tilde{\psi} - d_e^2 \tilde{\psi}'')&=&-ix (\tilde{\phi}+d_i \tilde{b}),  \label{lin1}\\
g(\tilde{\phi}'' -k^2 \tilde{\phi})&=&-ix(\tilde{\psi}'' -k^2 \tilde{\psi}),\\
g((1+k^2 d_e^2 )\tilde{b} -d_e^2 \tilde{b}'')&=&ix d_i (\tilde{\psi}'' -k^2 \tilde{\psi})+ix \tilde{v},  \label{lin4}\\
g\tilde{v}&=&ix \tilde{b},
\een
where $g=\gamma/(2k)$ and the prime symbol denotes derivative with respect to the argument.  We remark then that the linear  system (\ref{lin1})--(\ref{lin4}) corresponds also to the linearization of the four-field model studied in Ref.~\onlinecite{Fit04}, provided
one uses $\psi_{eq}=-x^2$ instead of $\psi_{eq}=-x^2/2$ and replaces the constant $d_{\beta}$ of Ref.~\onlinecite{Fit04} with the constant $d_i$ (which corresponds to taking the limit $\beta \rightarrow +\infty$).  The additional terms of the model of (\ref{flux})--(\ref{v}) that are absent in the four-field model of Ref.~\onlinecite{Fit04}, indeed,  contribute only during the nonlinear regime.   We can then export the analysis carried out in Ref.~\onlinecite{Fit04}, where a relation for  the growth rate in terms of equilibrium parameters was found by  asymptotic matching. In this way we obtain the following expression for  the growth rate for our  system of   (\ref{lin1})--(\ref{lin4}):
 \beq  \label{disprel}
 -\frac{\pi}{\Delta '}-\frac{\pi}{2} \frac{g^2}{d_i G(g/d_i)}+\frac{d_e d_i G(g/d_i)}{g}=0,
 \eeq
 where $\Delta '$ is the classical tearing stability parameter \cite{Fur63} and $G(x)=(\sqrt{x}/2)\Gamma(1/4+x/4)/\Gamma(3/4+x/4)$, with $\Gamma$ indicating the Gamma function.
 A version of the relation (\ref{disprel}), adopting resistivity instead of electron inertia, has also been used in Ref.~\onlinecite{Baa11} for linear studies of reconnection based on 2D incompressible Hall MHD.
The  relation (\ref{disprel}) is valid if the conditions  $d_e \ll g \ll d_i \ll 1  $ are satisfied.\cite{Fit04}

\begin{figure}[h!]
\centering
\includegraphics[width=9cm]{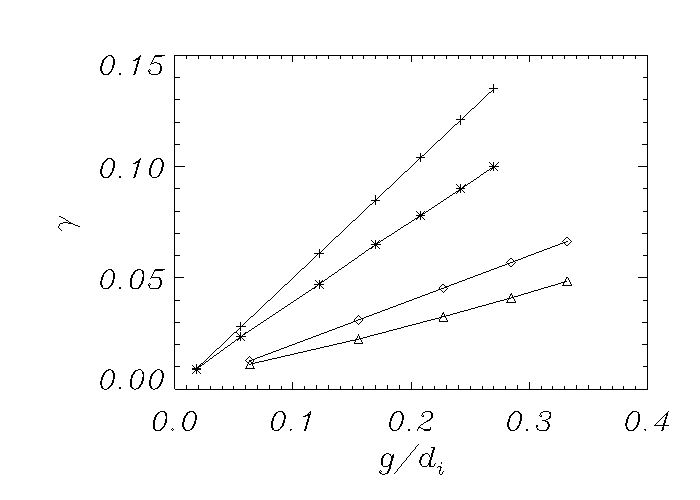}
\caption{Comparison between values of the growth rate $\gamma$ obtained from numerical simulations and from the asymptotic relation (\ref{disprel}), for different values of the parameter $g/d_i$. Crosses and asterisks indicate analytical and numerical values, respectively, for $d_i=0.5$, whereas diamonds and triangles correspond to analytical and numerical values, respectively, for $d_i=0.2$. For all cases $\Delta '= 59.9$.}
 \label{fig1}
\end{figure}

In Fig. \ref{fig1}, values of the growth rate $\gamma$, obtained from the asymptotic relation (\ref{disprel}), are checked against values obtained from numerical simulations, for a case with large $\Delta '$. The numerical code used for these  simulations is an adaptation  of the one used in Ref.~\onlinecite{Grasso07}  to solve the four-field system initialized  by perturbing about the equilibrium  
\beq   \label{eqnum}
\psi_{eq}=\frac{1}{\cosh^2 x}, \qquad b_{eq}= \phi_{eq}= v_{eq}=0.
\eeq
Note that the equilibrium (\ref{eqnum}), when expanded about $x=0$, corresponds to the equilibrium (\ref{eqlin}) adopted for deriving the analytical expression for the growth rate.

The model equations are solved on a grid consisting of up to 2048$\times$4096 grid points, depending on  the scale lengths to be resolved. All the fields are split into  the time-independent equilibrium and an evolving perturbation advanced in time by  a third order Adams-Bashforth algorithm. Periodic boundary conditions have been imposed along the shear equilibrium magnetic field direction, $y$, whereas Dirichlet conditions have been applied in the  $x$-direction with  all the perturbed fields vanishing at the boundaries. A pseudospectral method is adopted for the periodic direction, while a compact finite difference algorithm on a non-equispaced grid is used for the spatial operations along the $x$ direction. The tearing instability is initiated by perturbing the equilibrium with a small disturbance of the parallel current density $j = - \lap \psi$ of the form $\delta j\left( {x,y} \right) = \delta j\left( x \right)\cos ({{2\pi y}/{L_y}})$, where $\delta j\left( x \right)$ is a function localized within a width of  order $d_e$ around the rational surface $x=0$. 

  One can observe from the figure that the agreement between numerical and analytical values becomes better and better as the parameter $g/d_i$ decreases. This is expected, since, the relation (\ref{disprel}) holds in the asymptotic limit $g/d_i \ll 1$. We remark that, in the large $\Delta '$ regime, in the limit $g/d_i \ll 1$, the relation (\ref{disprel}) can be approximated by \cite{Fit04}
\beq
g=\frac{1}{\sqrt{2 \pi}} \frac{\Gamma(1/4)}{\Gamma(3/4)} \sqrt{d_e d_i},
\eeq
which also shows the accelerating role played by the Hall term, associated with the length  scale $d_i$.

%%%%%%%%%%%%%%%%%%%%%%%%%%%%%%%
%%%%%%%%%%%%%%%%%%%%%%%%%%%%%%%
\subsection{Nonlinear numerical simulations}

Having obtained a handle on the linear dynamics, we now describe  our  nonlinear numerical simulations.  In particular, we  follow the nonlinear evolution of the velocity and magnetic fields during the  process of magnetic reconnection initiated by perturbing the equilibrium (\ref{eqnum}).  The code employed is that of Sec.~\ref{sec:rlin}.

\begin{figure*}[htb]
 \centering
 \subfigure[]{
\includegraphics[width=0.4\textwidth]{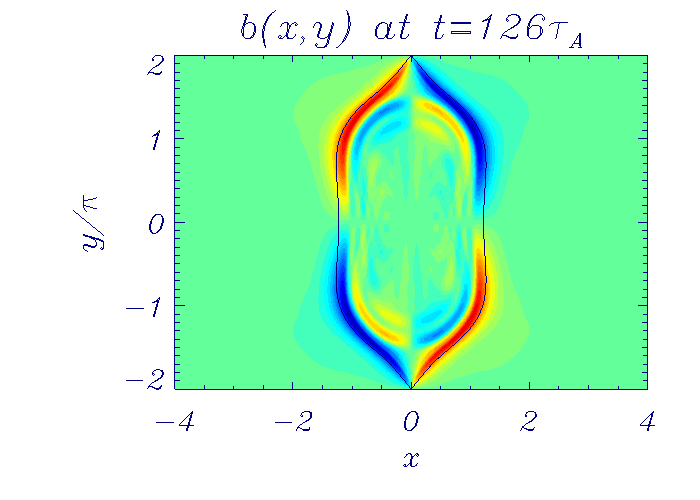}
   \label{2a}
 }  
 \subfigure[]{
\includegraphics[width=0.4\textwidth]{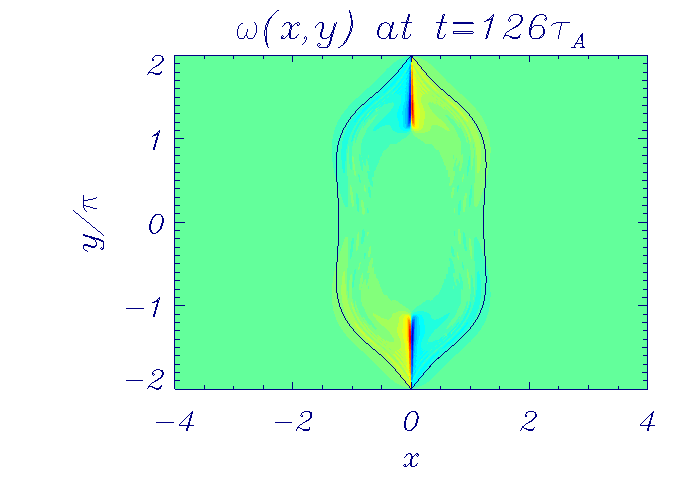}
   \label{2c}
    } 
     \subfigure[]{
\includegraphics[width=0.4\textwidth]{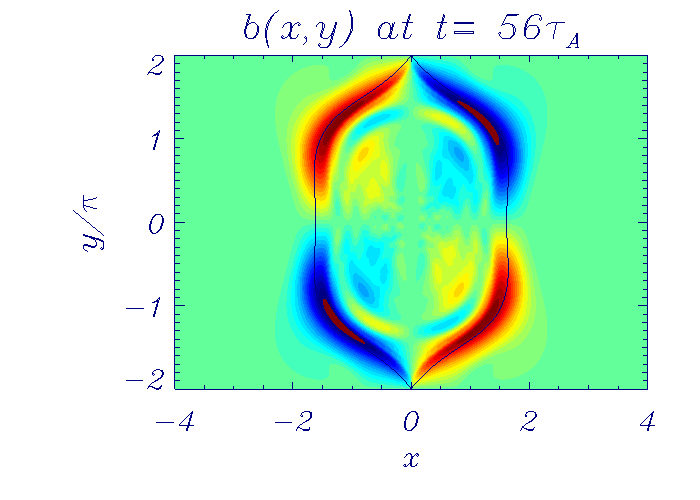}
   \label{2b}
    }
 \subfigure[]{
\includegraphics[width=0.4\textwidth]{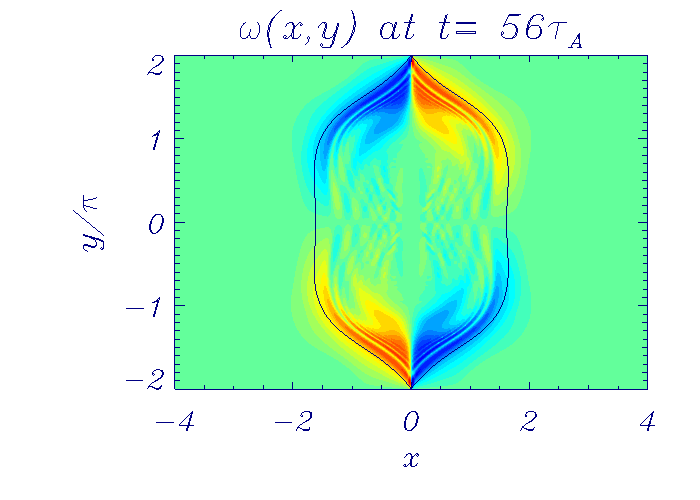}
   \label{2d}
 } 
 \caption{Contour plots of the out-of-plane magnetic field $b$ and vorticity  $\omega$ at time $126\tau_A$ for $d_e=0.05$ and $d_i=0.5$ (Figs.~\ref{2a} and \ref{2c}) and at time $56\tau_A$ for $d_e=0.1$ and $d_i=1$ (Figs.~\ref{2b} and \ref{2d}) . The magnetic island is superimposed on  the contour plots.}
 \label{fig2}
\end{figure*}

\begin{figure*}[htb]
 \centering
 \subfigure[]{
\includegraphics[width=0.4\textwidth]{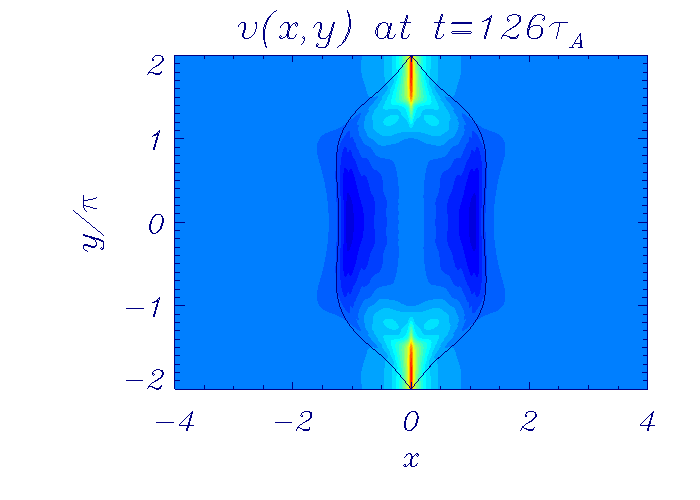}
   \label{v126}
 }  
 \subfigure[]{
\includegraphics[width=0.4\textwidth]{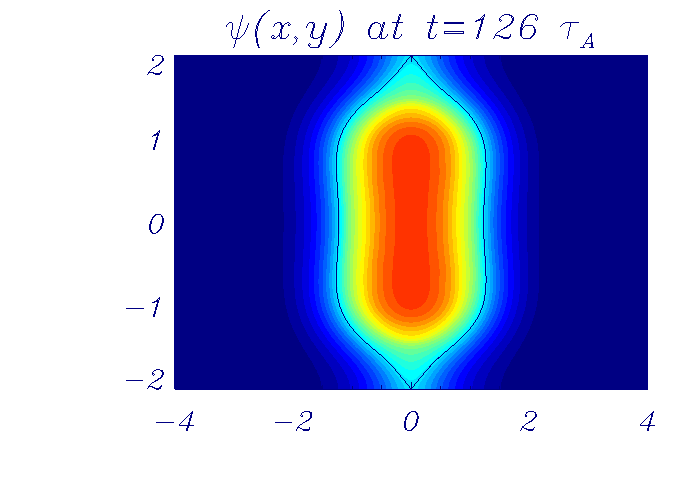}
   \label{psi126}
 } 
 \caption{Contour plots of the velocity field $v$ Fig.~\ref{v126} and flux  $\psi$ Fig.~\ref{psi126} at time $126\tau_A$ for $d_e=0.05$ and $d_i=0.5$. The magnetic island is superimposed on  the contour plots.}
 \label{vpsi126}
\end{figure*}

Figure \ref{fig2} shows  contour plots of the out-of-plane magnetic and vorticity fields, at times well into the nonlinear regime, for two choices of  skin depths  with the same mass ratios.  
The two times, $126\tau_A$ for the case with $d_e=0.05, d_i=0.5$ and  $56\tau_A$ for the case with $d_e=0.1, d_i=1$, were chosen because they represent approximately the same nonlinear stage.   In both cases the field $b$ exhibits the characteristic quadrupolar structure,   a signature of Hall reconnection (see, e.g., Refs.~\onlinecite{Yam10,Uzd06}). In the case with the smaller skin depths,  Figs.~\ref{2a} and \ref{2c},  one observes that the vorticity concentrates on a narrow region with a size of  order of $d_e$. In this region the behavior is mainly dictated by incompressible hydrodynamics and the system can eventually become prone to the Kelvin-Helmholtz instability.\cite{Bis97}  On the other hand, when increasing $d_e$ and $d_i$, as in  Figs.~\ref{2b} and \ref{2d}, vorticity is no longer concentrated on a narrow region but distributes mainly along the island separatrices and inside the island over a region of width on the order of $ d_i$, thus suppressing the Kelvin-Helmholtz instability. A similar mechanism for inhibiting a secondary Kelvin-Helmholtz instability was observed also for collisionless reconnection in the presence of a guide field in Refs.~\onlinecite{Del06,Gra09,Tas10}. In this case, the role of the Hall term was played by the  electron pressure  contribution to  Ohm's law.  For completeness, we plot the remaining two fields at $126\tau_A$ in Fig.~\ref{vpsi126} with $v$ shown in Fig.~\ref{v126} and $\psi$ in Fig.~\ref{psi126}.

\begin{figure*}[htb]
 \centering
 \subfigure[]{
\includegraphics[width=0.4\textwidth]{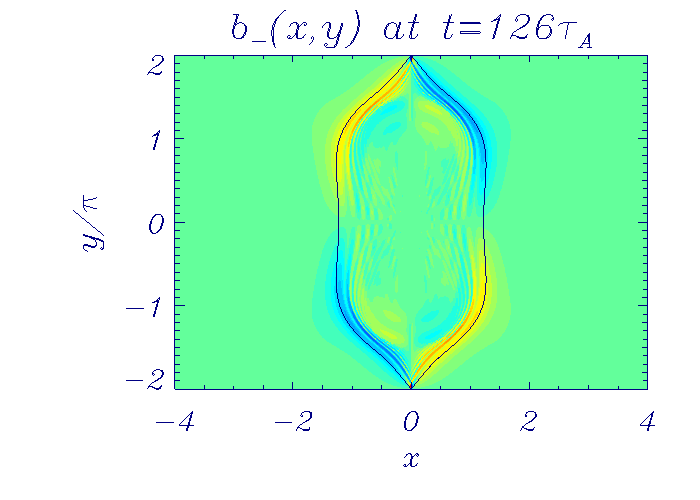}
   \label{b-}
 }  
 \subfigure[]{
\includegraphics[width=0.4\textwidth]{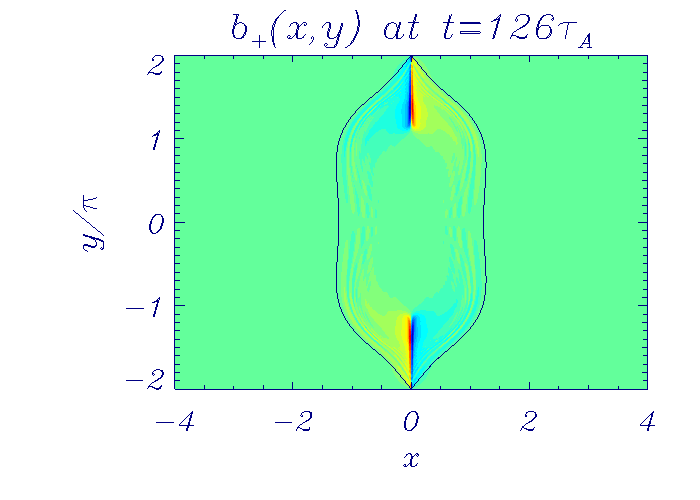}
   \label{b+}
    } 
     \subfigure[]{
\includegraphics[width=0.4\textwidth]{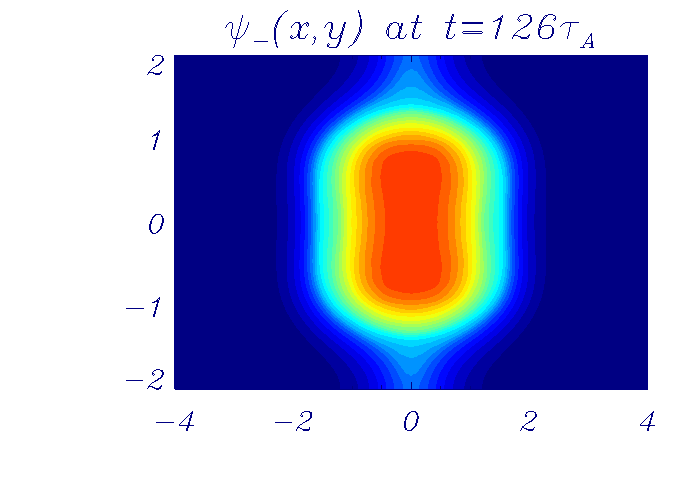}
   \label{psi-}
    }
 \subfigure[]{
\includegraphics[width=0.4\textwidth]{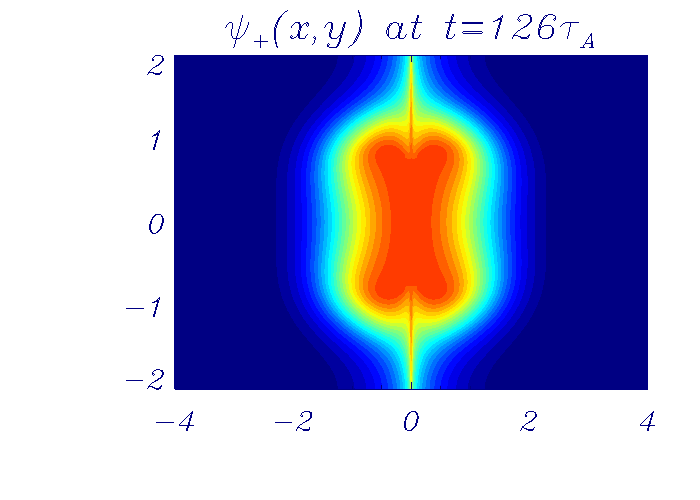}
   \label{psi+}
 } 
 \caption{Contour plots of normal fields $(b_{\pm},\psi_{\pm})$ at time $126\tau_A$ for $d_e=0.05$ and $d_i=0.5$. The magnetic island is superimposed on  the contour plots.}
  \label{normalF}
\end{figure*}

It is of interest to compare the four fields $(\omega, b, v,\psi)$ with the normal fields $(b_{\pm},\psi_{\pm})$  This is done in  Fig.~\ref{normalF} for the smaller skin depths.    As noted above in Figs.~\ref{2a} and  \ref{2c}, $b$ and $\omega$ display the characteristic quadrupolar  and current layer behavior, respectively, while from Figs.~\ref{v126} and \ref{psi126} the field $v$ is seen to display a sort of amorphous structure with a mixture of  both features,  while  $\psi$ shows an elongated form with minimal current layer evidence.  In comparison the normal fields of Fig.~\ref{normalF} reveal a cleaner separation of behavior, with $(b_+,\psi_+)$  displaying the current layer, which is notably absent in the normal fields $(b_-,\psi_-)$.  Observe the amorphous behavior of $v$ is absent and the quadrupolar behavior of $b$ has now been concentrated along the magnetic island contour.  We draw the conclusion that the normal fields more clearly delineate the nature of the evolution. 
 
\begin{figure} 
\centering
\includegraphics[width=8cm]{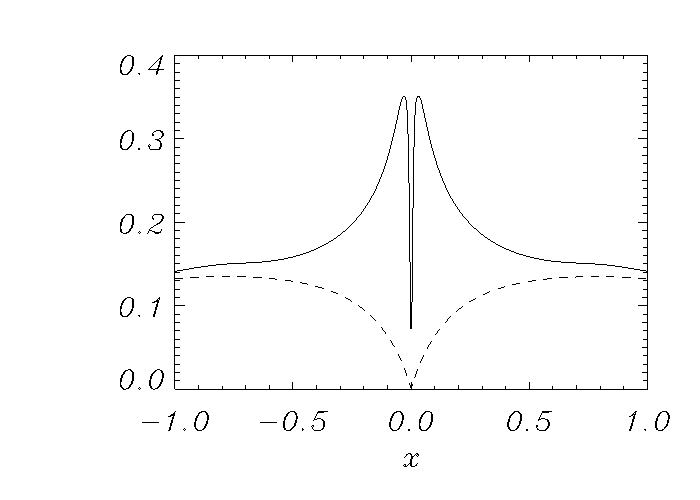}
\caption{Plots of $\vert \nabla \phi_+\vert$ (dashed line) and $\vert \nabla \phi_- \vert$ (solid line) at $y=0$ and $t=126$.} 
 \label{fig3}
\end{figure}

As noted in Sec.~\ref{ssec:normalF}, it is evident from Eqs.~\eqref{n1} that the normal fields $\psi_{\pm}$ correspond to the $z$-components of the electron and ion canonical momenta.  It is important to recall that the notion of canonical momentum originates in the Hamiltonian formalism and, consequently, that they should play a clarifying role in the present application is not surprising.  Because  the fields $\psi_{\pm}$  are advected by the velocities associated with $\phi_{\pm}$ (cf.\  Eqs.~(\ref{psipm})), we examine  the moduli  $\vert \nabla \phi_+\vert$ and $\vert \nabla \phi_- \vert$ of the perpendicular velocities $\mathbf{V}_{\pm}=\widehat{z}\times\nabla\phi_{\pm}$  that are doing the advecting.   Figure \ref{fig3} shows profiles of the moduli  at $y=0$, i.e. across the X-point.   One observes that, in a narrow region around the resonant surface at $x=0$, the velocity 
$\mathbf{V}_-$, which is predominantly due to the electrons, dominates over the  the velocity 
$\mathbf{V}_+$, which is predominantly due  the ions  and actually vanishes at $x=0$. This is consistent with  the behavior described in Ref.~\onlinecite{Bis97} where  magnetic flux bundle coalescence in a high $\beta$ regime was investigated. As discussed in Ref. \onlinecite{Bis97}, this behavior can be explained considering that, in a region with the size $L \sim d_e \ll d_i$ around the resonant surface, the system (\ref{flux})--(\ref{v}) reduces to 2D electron MHD. The dynamics is then essentially governed by electron motion, whereas ions are immobile. On the other hand, on scales  $L \geq d_i$, one enters an MHD regime, where $v_{e \perp} \approx v_{i \perp}$, as Fig.~\ref{fig3} shows with corrections due to the use of the normal fields. 

\begin{figure*}[htb]
 \centering
 \subfigure[]{
\includegraphics[width=0.4\textwidth]{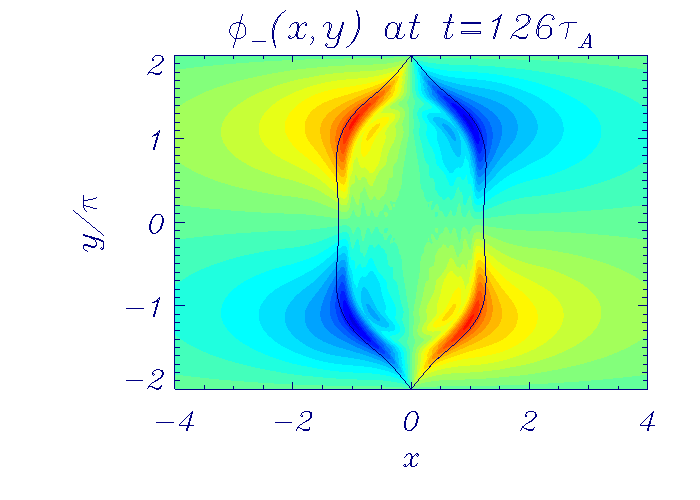}
    } 
 \subfigure[]{
\includegraphics[width=0.4\textwidth]{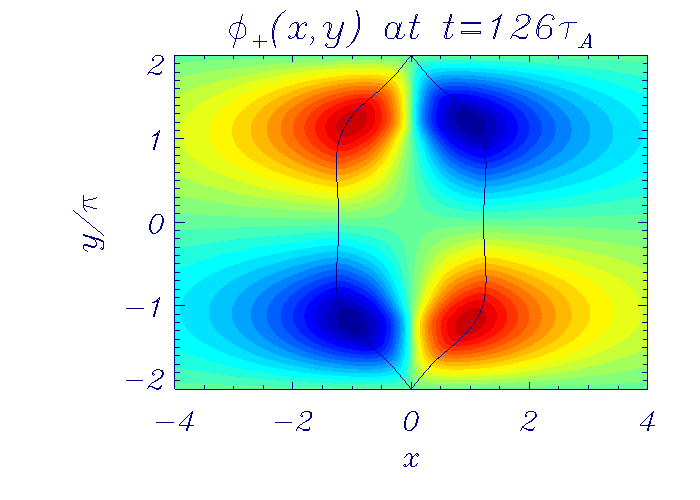}
       } 
 \subfigure[]{
\includegraphics[width=0.4\textwidth]{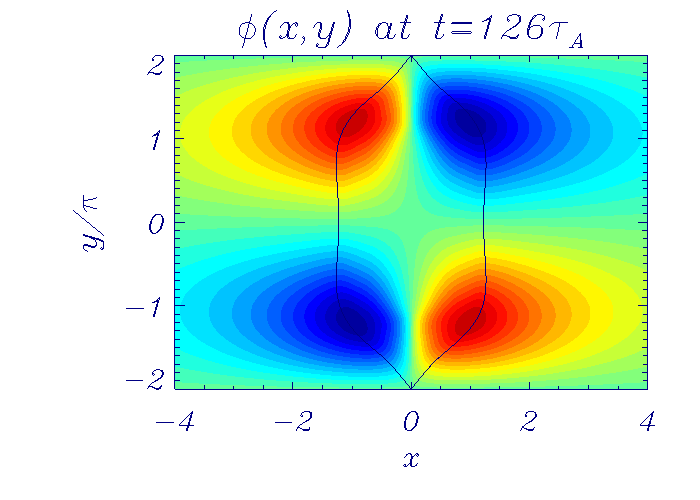}
    } 
 \caption{Contour plots of $\phi$ (left), $\phi_+$ (middle) and $\phi_-$ (right) for $d_e=0.05$ and $d_i=0.5$. The magnetic island is superimposed to the contour plots..}
 \label{fig4}
\end{figure*}

The global structures of the velocities   $\mathbf{V}_{\pm}$  are  revealed in the  contour plots of $\phi_+$ and $\phi_-$, respectively, shown in Fig. \ref{fig4}.    As expected, upon comparing $\phi$ to $\phi_+$ it is seen that the bulk velocity is mostly due to  the ion velocity, which exhibits the characteristic convective cells.   The stream function $\phi_-$, associated with the corrected electron velocity, on the other hand, concentrates mainly in narrow structures along the separatrices. An analytical argument justifying such behavior of the electron velocity was provided in Ref.~\onlinecite{Bis97}, based on the electron MHD approximation, valid on scales much smaller than $d_i$.

 %%%%%%%%%%%%%%%%  snip  %%%%%%%%%%%%%%%%%%%%%%
 
\begin{figure} 
\centering
\includegraphics[width=9.5cm]{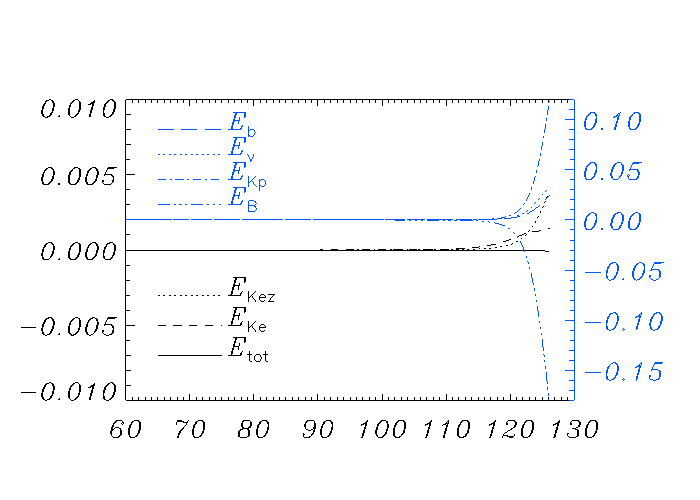}
\caption{Temporal plots of the terms of the Hamiltonian ${\mathscr{H}}$ of \eqref{sumE} relative  to their values  at $t=0$ for parameters values  $d_e =0.05$ and $d_i=0.5$.  The total energy, represented by $E_{\rm tot}$,  retains its initial value  (solid line)  throughout the  simulation.  The other terms, defined by \eqref{Kp}-\eqref{Kez},  are seen to increase at the expense of the decreasing perpendicular magnetic energy $E_{\rm B}$ (dash-dot-dot-dot) with most energy going into the perpendicular kinetic energy $E_{\rm Kp}$ (dash-dot).  Note that times before $t=60\tau_A$ are not shown  because the dynamics is  still in  the linear phase where variations of all  terms are  negligible.}
 \label{energy}
\end{figure}

Because of the Hamiltonian nature of the model the total energy of \eqref{RE}  is conserved, yet during the course of the dynamics energy may transfer  from one term to another.  In order to track this  we write
\begin{equation}
\mathscr{H}=  \mathscr{H}_{\rm Kp} +\mathscr{H}_{\rm v} +\mathscr{H}_{\rm B} +\mathscr{H}_{\rm b} +\mathscr{H}_{\rm Ke}  +\mathscr{H}_{\rm Kez} 
\label{sumE}
\end{equation}
with
\ben
\label{Kp}
\mathscr{H}_{\rm Kp} &=&\int\! d^2x\,   |\nabla \phi |^2/2
\\
\mathscr{H}_{\rm v} &=&\int\! d^2x\,  v^2/2
\\
\mathscr{H}_{\rm B} &=&\int\! d^2x\,  |\nabla \psi |^2/2
\\
\mathscr{H}_{\rm b} &=&\int\! d^2x\,  b^2/2
\\
\mathscr{H}_{\rm Ke}&=&\int\! d^2x\,   d_e^2\, |\nabla^2 \psi|^2/2
\\
\mathscr{H}_{\rm Kez}&=&\int\! d^2x\,    d_e^2 \, |\nabla b|^2/2
\label{Kez}
\een
and track each term during the reconnection process.   Here, for convenience and physical clarity,  we do this in terms of  the original  fields that appear in the Hamiltonian as a sum of squares,  rather than evaluate the expression of \eqref{normalH}  in terms of the normal fields.  In Fig.~\ref{energy} the total energy  $\mathscr{H}$ is displayed as a solid line, showing that indeed the numerics preserves it well up to time $126\tau_A$ for our example with $d_e =0.05$ and $d_i=0.5$.  Instead of plotting $\mathscr{H}$, we plot  $E_{\rm tot}$, which is the total energy relative to the initial value of $\mathscr{H}$.  The other energies of \eqref{Kp}-\eqref{Kez} are plotted similarly,  e.g. $E_{\rm Kp}$ is the value of  $\mathscr{H}_{\rm Kp}$, relative to  $\mathscr{H}$.  Next we observe that the energy $E_{\rm B}$ decreases while getting transferred  to all of the other terms in varying amounts.   The energy $E_{Ke}$, which is essentially the electron kinetic energy,  gains only a small amount, as does the energy $E_{\rm Kez}$, which also contains higher order derivatives.  Note both these energies are referred to the left hand scale.  All of the other energies grow significantly more, but by far most of the energy goes into $E_{\rm Kp}$, the perpendicular kinetic energy, which significantly dominates $E_{\rm b}$, the parallel magnetic,  and  $E_{\rm v}$  parallel kinetic energies.

 To compare these results with more conventional analyses we consider the approximate Hamiltonian  of (\ref{happr}), which although not exactly conserved can be used  to proved  a physically transparent interpretation of the energy redistribution process during the reconnection. We write the approximate  Hamiltonian $\tilde{\mathscr{H}}$  as the sum 
\beq
\tilde{\mathscr{H}}=\tilde{\mathscr{H}}_{\rm B} + \tilde{\mathscr{H}}_{\rm i} + \tilde{\mathscr{H}}_{\rm e},
\eeq
where $\tilde{\mathscr{H}}_{\rm B}=(1/2)\int d^2x (\vert \nabla \psi \vert^2 +b^2)$ is the total magnetic energy, $\tilde{\mathscr{H}}_{\rm i}= (1/2)\int d^2x (v_{i \perp}^2 + v_{iz}^2)$ is the total  ion kinetic energy and 
$\tilde{\mathscr{H}}_{\rm e}=(1/2) (d_e^2 /d_i^2)\int d^2 x\left(v_{e \perp}^2 +v_{ez}^2\right)$ is the total electron kinetic energy.   It is easy to infer from  Fig.~\ref{energy}  that reconnection converts most of the magnetic energy into kinetic energy of the ion flow. This is consistent with what is heuristically mentioned in Ref.~\onlinecite{Bis97}, although the actual conserved energy was not identified in that reference.

%%%%%%%%%%  snip %%%%%%%%%%%%%%%%%

%%%%%%%%%%%%%%%%%%%%%%%%%%%%%%%
%%%%%%%%%%%%%%%%%%%%%%%%%%%%%%%
%%%%%%%%%%%%%%%%%%%%%%%%%%%%%%%
\section{Summary and Conclusions}
\label{sec:conclu}

In this paper we have given a comprehensive analysis of reduced extended MHD, a 2D version of extended MHD.  We have derived the model starting from the Hamiltonian form of L\"ust's equations by a reduction procedure that produced the Hamiltonian form of the reduced model, RXMHD.   This procedure led to the physical energy, which serves as the reduced Hamiltonian,  the four families of Casimir invariants, and the definitions of the normal fields $(b_{\pm},\psi_{\pm})$ in terms of which the four equations of motion take a simplified intuitive form.  Further reductions of the RXMHD led in a natural way to reduced  Hall MHD,  inertial MHD, and ideal MHD, Hamiltonian field theories with conserved energies and associated Casimir invariants.  Analyses of RXMHD revealed the natural modes of oscillation,   the expected whistler and  ion cyclotron waves.  The analytical expression for the linear collisionless  tearing growth rate was inferred and checked against numerical solutions.  Nonlinear simulations of collisionless  tearing revealed behavior typical of Hall and electron inertia physics, but better organized by the new normal field variables. 

The content of this work opens many avenues for further study, both analytical and numerical.  We mention a few. On the analytical side one can effect absolute equilibrium calculations akin to those of Refs.~\onlinecite{HLS16, MLM16} in order to infer energy cascades. In addition one can derive the Hamiltonian form of 3D incompressible XMHD using the Dirac constraint technique of Ref.~\onlinecite{Cha12} and derive the weakly 3D version of the present model, where the latter  gives rise to terms linear in parallel derivatives caused by a strong guide field.   The general Hamiltonian form of such weakly 3D models is available in Ref.~\onlinecite{Tas10} and the correct one can be obtained by aspect ratio expansion  of the full XMHD model or by Hamiltonian reduction.  Having in hand the weakly 3D version of RXMHD opens the way for  numerical treatment of weakly 3D collisionless tearing.

%%%%%%%%%%%%%%%%%%%%%%%%%%%%%%%
%%%%%%%%%%%%%%%%%%%%%%%%%%%%%%%
%%%%%%%%%%%%%%%%%%%%%%%%%%%%%%%

\section*{Acknowledgments}

\noindent ET was supported by the CNRS by means of the PICS project NEICMAR. HMA would like to thank the Egyptian Ministry of Higher Education for supporting his research activities. PJM was supported by U.S. Dept.\ of Energy  under contract \#DE-FG02-04ER-54742.  He would also like to acknowledge support from the Alexander von Humboldt Foundation and the hospitality of the Numerical Plasma Physics Division of the IPP, Max Planck, Garching.


\begin{thebibliography}{10}

\bibitem{rmhd} H. R. Strauss,  Phys. Fluids {\bf19}, 134 (1976). 

\bibitem{HKM85}  R. D. Hazeltine, M. Kotschenreuther, and P. J. Morrison, Phys. Fluids {\bf 28}, 2466 (1985).

\bibitem{DA84} J. F. Drake and T. M. Antonsen, Jr., Phys. Fluids {\bf 27}, 898 (1984).

\bibitem{HW83} A. Hasegawa and M. Wakatani, Phys. Fluids {\bf26}, 1770 (1983).
  
\bibitem{Bri92} A. Brizard, Phys. Fluids B {\bf 4}, 1213 (1992).

\bibitem{Dor93} W. Dorland, G.W. Hammett, Phys. Fluids B {\bf 5}, 812 (1993).

\bibitem{Sny01} P.B. Snyder, G.W. Hammett, Phys. Plasmas {\bf 8}, 3199 (2001).

\bibitem{SSB05} D. Strintzi, B. D. Scott, and A. J. Brizard, Phys. Plasmas {\bf12}, 052517 (2005).

\bibitem{SPK94} T. J. Schep, F. Pegoraro, and B. N. Kuvshinov, Phys. Plasmas 

\bibitem{GCPP01}  D.  Grasso,   F. Califano, F. Pegoraro,  and F. Porcelli,   
 Phys. Rev. Lett. {\bf 86}, 5051 (2001). 

\bibitem{Grasso07} D. Grasso, D. Borgogno and F. Pegoraro, Phys. Plasmas  {\bf 14}, 055703 (2007).

\bibitem{TMWG08}   E. Tassi,  P. J. Morrison,  F. L. Waelbroeck,  D. Grasso, Plasma Phys. and Control. Fusion {\bf 50}, 085014 (2008).

\bibitem{WHM09} F. L. Waelbroeck, R. D. Hazeltine, and P. J. Morrison Phys. Plasmas {\bf16}, 032109  (2009).

\bibitem{Sco10} B. Scott, Phys. Plasmas {\bf 17}, 102306 (2010).

\bibitem{Wae12} F.L. Waelbroeck and E. Tassi, Commun. Nonlinear Sci. Numer. Simulat. {\bf 17}, 2171 (2012).

\bibitem{KWM15} I. Keramidas Charidakos, F. Waelbroeck, and P. J. Morrison, 
 Physics of Plasmas {\bf22}, 112113 (2015).

\bibitem{MG80} P. J. Morrison and J. M. Greene, Phys. Rev. Lett.  {\bf 45}, 790 (1980).

\bibitem{M98} P.J. Morrison, Rev. Mod. Phys. {\bf70}, 467 (1998).

\bibitem{M05} P. J. Morrison, Phys.Plasmas {\bf 12}, 058102 (2005).

\bibitem{MH84} P. J. Morrison and R.D. Hazeltine {\bf27}, 886 (1984). 

\bibitem{MM84} J. E. Marsden and P. J. Morrison, Contemp.\ Math. {\bf28} 133 (1984) 

\bibitem{HHM86} C. T. Hsu, R. D. Hazeltine, and P. J. Morrison, 
Phys. Fluids {\bf29}, 1480 (1986).

\bibitem{Haz87}  R. D. Hazeltine, C. T. Hsu and P. J. Morrison, Phys. Fluids {\bf 30}, 3204 (1987).

\bibitem{Lust} R.  L\"ust, Fortschritte der Physik {\bf7}, 503 (1959).

\bibitem{Abd15} H. M. Abdelhamid, Y. Kawazura and Z. Yoshida,   J. Phys.  A  {\bf 48},  235502 (2015).
 
\bibitem{Lin15}  M. Lingam, P. J. Morrison,  and G. Miloshevich,  Phys. Plasmas {\bf 22}, 072111 (2015).

\bibitem{Bis97} D. Biskamp, E. Schwarz, J. F. Drake, Phys. Plasmas {\bf 4}, 1002 (1997).

\bibitem{Del06} D. Del Sarto, F. Califano, F. Pegoraro, Mod. Phys. Lett. B {\bf 20}, 931 (2006).

\bibitem{Gra09} D. Grasso, D. Borgogno, F. Pegoraro, E. Tassi, Nonlin. Processes Geophys. {\bf 16}, 241 (2009).

\bibitem{Tas10} E. Tassi, P. J. Morrison, D. Grasso, F. Pegoraro, Nucl. Fusion {\bf 50}, 034007 (2010).

\bibitem{Com12} L. Comisso, D. Grasso, E. Tassi and F.L. Waelbroeck, Phys. Plasmas, {\bf 19}, 042103 (2012).

\bibitem{And14} N. Andres, L. Martin, P. Dmitruk and D. P. Gomez, Phys. Plasmas, {\bf 21}, 072904 (2014).

\bibitem{Kimura14} K. Kimura and P. J. Morrison, Phys. Plasma {\bf 21}, 082101 (2014).

\bibitem{amp0}  T. Andreussi, P. J. Morrison, and F. Pegoraro, Plasma Phys.  and Control. Fusion {\bf 52}, 055001 (2010).

\bibitem{amp1}T. Andreussi, P. J. Morrison, and F. Pegoraro,  Phys. Plasmas {\bf 19}, 052102  (2012).

\bibitem{Thi00} J. L. Thiffeault and P. J. Morrison, Physica D {\bf 136}, 205 (2000).
  
\bibitem{Hamdi16} H. M. Abdelhamid and Z. Yoshida, Phys. Plasmas {\bf 23}, 022105 (2016).

\bibitem{Fit04} R. Fitzpatrick and F. Porcelli, Phys. Plasmas  {\bf 11}, 4713 (2004).
 
\bibitem{Fur63} H. P. Furth, J. Killeen and M. N. Rosenbluth, Phys. Fluids  {\bf 6}, 459 (1963). 

\bibitem{Baa11} S. D. Baalrud, A. Bhattacharjee, Y.-M. Huang and K. Germaschewski, Phys. Plasmas {\bf 18}, 092108 (2011).

\bibitem{Yam10} M. Yamada, R. Kulsrud, H. Ji, Rev. Mod. Phys.  {\bf 82}, 603 (2010).

\bibitem{Uzd06} D. A. Uzdensky, R. M. Kulsrud, Phys. Plasmas  {\bf 13}, 062305 (2006).

\bibitem{HLS16} H. M. Abdelhamid, M. Lingam and S. M. Mahajan, Astrophys. J. {\bf 829}, 87 (2016).

\bibitem{MLM16} G. Miloshevich, M. Lingam, and P. J. Morrison. {\it On the structure and statistical theory of turbulence of extended magnetohydrodynamics},   arXiv:1610.04952v1 [physics.plasm-ph] (2016). 

\bibitem{Cha12} C. Chandre, P. J. Morrison and E. Tassi, Phys. Lett. A {\bf 376}, 737 (2012).



\end{thebibliography}
\end{document}